# Numerical Techniques for the Maximum Likelihood Toeplitz Covariance Matrix Estimation: Part I. Symmetric Toeplitz Matrices

Yuri Abramovich, *Life Fellow, IEEE*, Victor Abramovich, *Member, IEEE*, Tanit Pongsiri

***Abstract* – In several applications, one must estimate a real-valued (symmetric) Toeplitz covariance matrix, typically shifted by the conjugated diagonal matrices of phase progression and phase "calibration" errors. Unlike the Hermitian Toeplitz covariance matrices, these symmetric matrices have a unique potential capability of being estimated regardless of these beam-steering phase progression and/or phase "calibration" errors. This unique capability is the primary motivation of this paper.**

## I. Introduction

We consider the traditional problem of the maximum likelihood (ML) estimation of a covariance matrix that is known to be a symmetric Toeplitz matrix, [1]-[14] potentially multiplied (from both sides) on the direct and conjugated diagonal matrices of phase progression and phase "calibration" errors. In this paper, we do not investigate the antenna calibration problem. Instead, we intend to develop a Toeplitz matrix estimation technique that applies to the presence of these phase errors. Note that, the generic methodology of ULA array calibration for an arbitrary Hermitian Toeplitz matrix, introduced in [24] - [25], is applicable in this case. Yet, the potential capability of the Toeplitz matrix estimation, regardless of the presence of the beam-steering and/or "calibration" phase errors, exists only for the symmetric Toeplitz covariance matrices, and this capability is the primary motivation of this paper. Distinctions between the number of free parameters that describe symmetric and Hermitian Toeplitz matrices are the main reason for this important distinction. For this reason, some specific techniques applicable to symmetric Toeplitz matrices may not apply to the Hermitian Toeplitz matrix case.

In Sec. II, we introduce the problem of the maximum likelihood symmetric Toeplitz matrix estimation and the M.T. Chu theorem [26] that specifies the set of parameters uniquely describing this class of matrices. This theorem describes the property of the true symmetric matrices, and therefore, to use this theorem constructively, we had to propose a numerical procedure for the unique reconstruction of the symmetric Toeplitz covariance matrix given the set of parameters estimates as specified by the Chu theorem. After conversion of the traditional sample matrix into a positive definite (p.d.) Toeplitz symmetric matrix, based on the Chu theorem parametrization, we then move to the development of the computational techniques for the maximum likelihood reconstruction of the symmetric Toeplitz matrix. The need for this step is justified by these "invariants" concerning phase steering and error values, and this matrix reconstruction does not deliver the globally optimal maximum likelihood Toeplitz matrix estimates. Yet, this is expected for the estimation in the presence of unknown interfering parameters. That is why our next move is the ongoing processing for the global ML Toeplitz matrix estimation in the absence of phase errors and the entire sample covariance matrix made available for optimization.

For the known a priori absence of any phase errors in the antenna array, one may use for the Toeplitz matrix reconstruction the entire sample matrix $\hat{\mathbf{R}}_N$ rather than its elements' moduli and eigenvalues. Yet, in this study, we use the derived moduli and eigenvalues of the Toeplitz matrices as the initial solutions for the search for the global likelihood ratio (LR) maximum. Before application of the MATLAB fmincon routine for the final LR maximization, we introduce the linear programming (LP) routine that modifies the estimated Toeplitz matrix in an attempt to equalize the eigenvalues of the product of the inverted Toeplitz and direct sample matrices. While fmincon can operate directly with the "moduli and eigenvalues" solutions, this LP step may be final, delivered by the convex routine, if the global LR maximum is not required.

Correspondingly, in Sec. III, we describe the integer technique for the Toeplitz matrix restoration using estimates of the Toeplitz matrix elements' moduli and eigenvalues. Due to the non-optimum estimation of the matrix elements moduli, the restored Toeplitz matrix has negative eigenvalues. Therefore, in Sec. IV, we describe the linear programming routine that trims the moduli of the Toeplitz matrix to get a p.d. Toeplitz solution or p.d. Toeplitz solution with several equal minimum positive eigenvalues, specified by Minimum Description Length (MDL)/Akaike Information Criterion (AIC) criteria, applied to the sample covariance matrix. In the reconstruction of the p.d. symmetric Toeplitz matrix, only the moduli of the matrix elements and eigenvalues estimates produced by the sample covariance matrix are used. Therefore, the introduced unique procedure, based on the Chu theorem, could be used for Toeplitz covariance matrix estimation in the

Yuri Abramovich, Victor Abramovich, and Tanit Pongsiri are with WR Systems, Ltd., Fairfax, VA 22030, USA. Emails: yabramovich@wrsystems.com, vabramovich@wrsystems.com, tpongsiri@wrsystems.com

presence of a “beam-steering” phase progression and/or phase “calibration” error. Unfortunately, these techniques cannot deliver the global ML optimum, providing the covariance matrix estimations in the presence of non-estimated phase errors.

While the original part of the paper, devoted to the unique possibility of estimation of the symmetric Toeplitz covariance matrix, is over, in Sec. IV and Sec. V, we continued our development of the computational techniques, ultimately delivering the global ML optimum. In Sec. IV, we introduce a linear programming tool to improve the LR of the “moduli and eigenvalues” solution by equalizing eigenvalues of the product of the sample matrix and inverted optimized symmetric Toeplitz matrix. In Sec. V, we introduce the MATLAB fmincon routine used for the global ML solution. In Sec. VI, we provide the results of the Monte-Carlo simulations, initially for the unique “moduli and eigenvalues” technique, followed by the techniques that exploit the true sample covariance matrix in the search for the global LR extremum. In Sec. VII, we conclude our paper.

## II. Specific Properties of the Symmetric Toeplitz Matrices: M.T. Chu Theorem

The existence of special techniques different from the ones applied to Hermitian Toeplitz matrices could be attributed to the very different number of parameters that describe the real-valued and complex-valued Toeplitz matrices. Specifically, the set of ($2N$ - 1) positive-valued parameters describing all covariance lags of the symmetric Toeplitz matrix is equal to the number of “free” positive parameters that describe the covariance lags moduli and eigenvalues of a symmetric Toeplitz matrix. For a Hermitian matrix, the number of positive parameters that describe one central positive and ($N$ - 1) complex-valued lags is equal to ($4N$ - 3), which significantly exceeds the number of elements’ moduli and eigenvalues ($2N$ - 1). This trivial observation explains why $N$ positive moduli values of $N$ real-valued covariance lags and ($N$ - 1) eigenvalues may describe the symmetric Toeplitz matrix but not the Hermitian one.

The M. T. Chu theorem [26] suggests that for the ($N$ - 1) given eigenvalues of the symmetric Toeplitz and $N$ moduli values of the matrix’s eigenvalues, there are only two possible reconstructions of the given symmetric matrix. One is the actual Toeplitz matrix $\mathbf{T}_N$, and the other is shifted in the direction $\theta'$, where

$$\frac{2\pi d}{\lambda}\sin\theta' = \pi, \tag{1}$$

so that

$$(\mathbf{T}_N)' = \mathrm{D}(\pi)\mathbf{T}_N\mathrm{D}(\pi), \tag{2}$$

where

$$\mathrm{D}(\pi) = \mathrm{diag}\left[1, \mathrm{e}^{\mathrm{i}\pi}, \mathrm{e}^{\mathrm{i}2\pi}, \dots, e^{i(N-1)\pi}\right], \tag{3}$$

$d$ is the inter-element spacing of the uniform linear array (ULA), and $\lambda$ is the wavelength. From (1) - (3), it follows that if the ULA operates in an oversampled regime, when

$$\frac{d}{\lambda} < \frac{1}{2}, \tag{4}$$

then there is only a single symmetric Toeplitz covariance matrix that fits the given eigenspectrum and moduli of the matrix elements. Therefore, irrespective of the existing beam-steering phase progression, when

$$\mathbf{T}_N(\theta_o) = \mathrm{D}(\theta_o)\mathbf{T}_N\mathrm{D}^{\mathrm{H}}(\theta_o), \tag{5}$$

$$\begin{gathered}\mathrm{D}(\theta_o) = \mathrm{diag}[1, \exp\left(i\frac{2\pi d}{\lambda}\sin\theta_o\right), \dots, \\ \exp\left(i(N-1)\frac{2\pi d}{\lambda}\sin\theta_o\right)],\end{gathered} \tag{6}$$

and/or the presence of antenna phase “calibration” random errors, when

$$\mathbf{R}_N(\mathbf{\Omega}_{N-1}) = \mathrm{D}(\mathbf{\Omega}_{N-1})\mathbf{T}_N\mathrm{D}^{\mathrm{H}}(\mathbf{\Omega}_{N-1}), \tag{7}$$

$$\begin{gathered}\mathbf{D}(\mathbf{\Omega}_{N-1}) = \mathrm{diag}\,[1, \mathrm{e}^{\mathrm{i}\varphi_1}, \dots, e^{i\varphi_{N-1}}], \\ -\varphi_{max} < \varphi_n < +\varphi_{max},\end{gathered} \tag{8}$$

the Toeplitz symmetric matrix may be accurately reconstructed using the moduli and eigenvalues of the matrix. While this finding does not exist for Hermitian Toeplitz matrices, it is desirable for practical applications. In practical applications, instead of the true covariance matrices, we deal with sample Hermitian matrices, calculated using $T$ i.i.d. training samples:

$$\begin{gathered}\hat{\mathbf{R}}_N = \frac{1}{T}\sum_{t=1}^{T}\mathbf{X}_t\mathbf{X}_t^{\mathrm{H}}, \quad \mathbf{X}_t = \mathbf{D}(\theta_o + \mathbf{\Omega}_{N-1})\mathbf{T}_N^{\frac{1}{2}}\boldsymbol{\xi}_t, \\ \boldsymbol{\xi}_t \sim \mathbb{C}\mathcal{N}(0, \mathbf{I}_N).\end{gathered} \tag{9}$$

Therefore, if we use their estimates derived from these sample matrices instead of the true moduli and eigenvalues, we may reconstruct the symmetric Toeplitz matrix estimate if our estimation methodology is sufficiently accurate. This paper concentrates on the symmetric Toeplitz matrix reconstruction, provided the elements’ moduli and eigenvalues of the sample matrix. At the same time, the peculiarities of the antenna calibration should be investigated separately.

The important issue to remember is that in the presence of additional effects, such as phase “calibration” errors or expectation of their presence, the maximum likelihood of the true covariance matrix may not be reached, as it would in the case of only estimating the covariance matrix. Nevertheless, our main task is to use the Chu theorem to reconstruct the symmetric Toeplitz matrix, given the estimates of the covariance matrix elements' moduli and matrix eigenvalues.

Since no attempts to exploit the Chu theorem have been reported in signal processing literature, let us first validate it by reconstructing the Toeplitz matrix given its true eigenvalues and elements' moduli. In our analysis, we selected the symmetric Toeplitz matrix $\mathbf{T}_N$:

$$\mathbf{T}_N = \sigma_n \mathbf{I}_N + \text{sinc}\,(W_2), \tag{10}$$

with $N = 17, W_2 = 0.1,$ and $\sigma_n^2 = 10^{-2}$ and

$$\text{sinc}\,(W_2) = \left[\frac{\sin 2\pi W_2\,(k-l)}{\pi(k-l)}\right], \quad k,l = 1,\dots,N. \tag{11}$$

In TABLE I. , we provide the first column of this matrix and the matrix eigenvalues.

TABLE I.

| **sinc ($W_2$) (first column)** | **$\lambda$(sinc ($W_2$))** |
|---|---|
| 0.21 | 1.0097 |
| 0.1871 | 0.9991 |
| 0.1514 | 0.8763 |
| 0.1009 | 0.4573 |
| 0.0468 | 0.0996 |
| 0.0000 | 0.0175 |
| -0.0312 | 0.0104 |
| -0.0432 | 0.0100 |
| -0.0378 | 0.0100 |
| -0.0208 | 0.0100 |
| 0.0000 | 0.01 |
| 0.0170 | 0.01 |
| 0.0252 | 0.01 |
| 0.0233 | 0.01 |
| 0.0134 | 0.01 |
| 0.0000 | 0.01 |
| -0.0117 | 0.01 |

Starting from the all-positive initial Toeplitz matrix $\mathbf{T}_N^+$:

$$\mathbf{T}_N^+ = [t_o, |t_1|, \dots, |t_{N-1}|], \tag{12}$$

where $|t_n|, n = 1,\dots,N-1$ are the true moduli of the covariance matrix elements, we distribute the sign inversions using the simple "maximum element" algorithm. At each step of this algorithm, all unoccupied positions by the sign change of the positive covariance lag moduli are tested, and the position where the sign inversion leads to the best criterion result is accepted. In [21], this algorithm was successfully applied for noise mitigation at the output of the antenna array with the integer phase control of the antenna's phase shifters.

A more advanced integer optimization routine was not required since we accurately reconstructed the true symmetric covariance matrix in this and a few other similar examples. Therefore, the Chu theorem is proven to be constructive, at least for the formulated conditions. To what extent the same simplistic algorithm may be applied in practical applications with the estimated matrix elements moduli and eigenvalues is a different issue, as addressed below.

## III. Reconstruction of the Symmetric Toeplitz Matrix Given the Elements Moduli and Eigenvalues of the Sample p.d. Matrix $\widehat{\mathbf{R}}_N$

In practical applications, we are usually given the traditional sample matrix $\widehat{\mathbf{R}}_N$. To reconstruct the symmetric p.d. Toeplitz matrix kernel in (11), we must find the maximum likelihood joint estimates for the Toeplitz matrix moduli of its sub-diagonal lags and eigenvalues using $\widehat{\mathbf{R}}_N$. Since the ML-optimum joint estimates are not directly available, we adopt the sub-optimum estimates first and then try to improve the likelihood.

As a reminder, the calculated moduli of the elements of the Hermitian sample matrix and its eigenvalues are available for this estimation. But first, let us demonstrate the severity of this estimation problem. In this specific test, instead of the parameters (moduli and eigenvalues) of the true covariance matrix $\mathbf{T}_N$, we used the sample matrix $\widehat{\mathbf{R}}_N$, averaged over $T$ = 17,000 ($T = 10^3 N$) i.i.d. training samples.

By applying the moduli redundancy averaging and matrix' eigenvalues and using the above-mentioned "integer maximum element" optimization algorithm, we could reconstruct a p.d. Toeplitz symmetric matrix. Yet, instead of the LR = 0.98 produced by the true covariance matrix, the reconstructed symmetric Toeplitz matrix provided LR = 0.4. When the same approach was applied to realistic sample volume cases $T = (2-10)N$, not a single (!) positive definite solution existed amongst all 65,535 possible sign inversions constellations over the matrix sub- (and super-) diagonals.

This test clarified that the LR maximization procedure for small and modest sample support needs to be carefully optimized to reach relatively high likelihood ratios for the estimated symmetric Toeplitz matrices. Specifically, instead of the non-existent "optimal" estimator, we have to consider a sequence of algorithms, gradually improving the properties of the estimated symmetric covariance matrices. Let us introduce the proposed routines.

### *3.1. Estimation of the Matrix Elements Moduli*

In the absence of the ML optimal estimation algorithm, let us start from the "naïve" moduli estimates, derived by the direct averaging of the moduli of the sub-diagonal elements:

$$|t_n| = \frac{1}{N-n}\sum_{k=1}^{N-n}|\hat{r}_{k,k+n}|. \tag{13}$$

The non-optimum nature of this estimate was demonstrated by the absence of a single positive definite symmetric Toeplitz matrix, irrespective of the sign inversions over the diagonals of this Toeplitz matrix. The impossibility of constructing a p.d. symmetric Toeplitz matrix for a small and medium training sample volume should be considered.

### *3.2. Eigenvalues Estimation*

It is well-known that the finite i.i.d. training sample support $T$ of the sample matrix $\widehat{\mathbf{R}}_N$ leads to an increased dynamic range of the sample matrix eigenvalues compared with the eigenvalues of the true covariance matrix [18]. Moreover, the typical spatial covariance matrices, where adaptive/optimum processing can provide significant signal-to-noise ratio improvement, have several identical minimal eigenvalues equal to the external white noise power.

In examples (10) and (11), the true matrix $\mathbf{T}_N$ has eleven identical minimal eigenvalues equal to the white noise power (see TABLE I. ). Therefore, the number of "noise subspace" eigenvalues has to be estimated using the Minimum Description Length or Akaike Information criterion. According to the MDL criterion, the number of noise subspace eigenvalues is equal to:

$$k_{\text{noise}} = N - \arg\min_k \left( -\log\left( \frac{\prod_{j=k+1}^{N} \hat{\lambda}_j^{\frac{1}{N-k}}}{\frac{1}{N-k}\sum_{j=k+1}^{N}\lambda_j} \right)^{(N-k)T} + \frac{1}{2}k(2N-k)\log T \right), \tag{14}$$

while the AIC criterion provides the following number:

$$k_{\text{noise}} = N - \arg\min_k \left( -\log\left( \frac{\prod_{j=k+1}^{N} \hat{\lambda}_j^{\frac{1}{N-k}}}{\frac{1}{N-k}\sum_{j=k+1}^{N}\lambda_j} \right)^{(N-k)T} + 2k(2N-k) \right). \tag{15}$$

The specified number of noise subspace eigenvalues allows it to proceed to the Random Matrix Theory (RMT) approach, which modifies the sample eigenvalues, reducing their dynamic range.

According to Theorem 3 in [38], the following quantities $\gamma_n, n = 1, \dots, N$ are strongly ($N$, $T$) consistent:

$$\hat{\gamma}_n = \frac{T}{K_n}\sum(\hat{\lambda}_K - \mu_K), \tag{16}$$

where $K_n$ is the number of equal eigenvalues $\hat{\gamma}_n$, and $\lambda_K$ and $\hat{\gamma}_k, k = 1, \dots, N$, are the traditional and new eigenvalues with multiplicity $\varkappa_n$, and $\hat{\mu}$ values:

$$\hat{\mu}_1 < \hat{\mu}_2 < \cdots < \hat{\mu}_N\ , \tag{17}$$

are the real-valued solutions to the following equation:

$$\frac{1}{N}\sum_{n=1}^{N}\frac{\hat{\lambda}_n}{\hat{\lambda}_n - \mu} = \frac{T}{N}\ (> 1). \tag{18}$$

Note that in the selected true Toeplitz covariance matrix (10) -(11) eigenvalue $\lambda_7$ is very close to the eleven noise subspace eigenvalues $\lambda_7 - \lambda_{17}$, and therefore, depending on the number of i.i.d. training samples $T$, the number of equal eigenvalues estimated by MDL/AIC criteria may reach $\lambda_7 - \lambda_{17}$, i.e. eleven eigenvalues.

### *3.3 Numerical Techniques for Symmetric Toeplitz Matrix Reconstruction Given the Estimated Moduli and Eigenvalues*

Applying "redundancy averaging" to the moduli of the diagonals of the sample matrix $\widehat{\mathbf{R}}_N$ to specify the moduli of the reconstructed Toeplitz symmetric matrix is not an optimal moduli estimation procedure which often leads to several negative eigenvalues in the restored Toeplitz matrix by sign change distributions over the matrix diagonals. For the $N = 17$-element ULA and $T = 85$, not a single sign inversion distribution, out of a possible 65,535, led to a p.d. Toeplitz matrix.

This fact makes it clear that the optimum reconstruction of a symmetric Toeplitz matrix, constructed with the redundancy averaged moduli and RMT-modified eigenvalues, should involve both "trimming" the moduli in the sub- (and super-) diagonals of the symmetric Toeplitz matrix along with the optimum distribution of the sign inversions over the matrix sub- (and super-) diagonals. Therefore, the mixed "integer-non-integer" optimization should be applied to this matrix optimization whereby the moduli and signs of the matrix diagonals should be jointly optimized to approach the RMT-specified Toeplitz matrix eigenvalues. Several software products for solving these mixed "integer-non integer" optimization problems (MINLF, Hexaly, APOPT, Gekko, mindPy) may be tested in the future.

In this study, we first tested the sequential application of the integer optimization of the sign inversion over the matrix sub- (and super-) diagonals with the original moduli, trying to get the best approximation of the eigenvalues specified above despite some number of minimum eigenvalues remaining negative. Note that another "naïve" attempt to improve this solution by the diagonal loading that brings the minimal eigenvalue to the specified positive value leads to a poor likelihood ratio of the loaded solution, and while it was also tested, it is not recommended. Instead, after the integer optimization that left several small negative eigenvalues, we "trim" the moduli by applying the Linear Programming routine to convert the symmetric Toeplitz matrix into a positive definite one.

Let us specify that for the "integer" part of optimization, instead of the simplistic "maximum element" routine, we tested the more sophisticated "integer dynamic programming routine" developed in [22] for the non-uniform linear array geometry optimization. In each of the "$N$ branches" of this routine, we forcefully change the sign of one element and then distribute other sign inversions, looking for the best criterion gain among all possible sign change positions. This distribution rule is the same as the "maximum element" algorithm, which differs from "dynamic programming" by searching for the best position for the first sign change over the matrix sub-diagonals.

In "dynamic programming," the preemptive sign change inversion creates each of the N branches in one of the ($N$ - 1) vacant sub-diagonals. As a result, we get up to ($N$ - 1) different solutions, equal to the number of "branches," and then select the best "branch" to finish the optimization. After the distribution of a few numbers of sign changes in each branch until the criterion stops improving, redistribution may then be implemented. For example, the first distributed sign change may be "returned" to its positive value, and the search for the new position for this element may be renewed. The idea is that test positions with the rest positive may be worse than a different position with the number of already made sign changes.

Moreover, we may apply a different criterion in the following iterative optimization procedure. For example, if we used the $L_2$ distance between the specified and optimized eigenvalues during the initial stage, we could apply a more sensitive minimax criterion in the second stage. One advantage of these distributional algorithms is that the criterion of optimization may be arbitrary. The only restriction at this stage of the symmetric matrix reconstruction is that we will use only the moduli of the elements and eigenvalues of the sample matrix. Therefore, the likelihood ratio cannot be used within the optimization procedure, while to what extent the available criteria on the eigenvalues of the matrix are correlated with the likelihood ratio is the most essential question that strongly affects the choice of the "working" criterion. In Sec. VI, where we introduce the optimization results, we specify these issues numerically.

Note that at this stage of the integer optimization with the remaining negative minimal eigenvalues, applying the optimized symmetric Toeplitz matrix for interference mitigation, for example, is impossible. Therefore, the following "layer" of complication would be achieved if the selection is conducted over the matrices that underwent linear programming conversion into a p.d. Toeplitz matrix. For example, selecting the best ($N$ - 1) branch solutions may be performed after the conversion of each solution to a p.d. symmetric Toeplitz matrix. Ultimately, the conversion to a positive definite Toeplitz matrix may be performed for each tested sign change position so that comparisons are conducted for p.d. Toeplitz matrices rather than for matrices with negative eigenvalues, as with the integer-only optimization.

This approach represents the above example of the "mixed integer-non-integer" optimization. Correspondingly, it is more computationally involved than the sequential integer and linear programming optimization of the sign changes and moduli trimming. Moreover, conversion to the p.d. Toeplitz matrices at every step of the sign inversion testing allows for applying a different and potentially more sensitive optimization criterion.

In this section, we introduced quite a broad spectrum of techniques for symmetric Toeplitz covariance matrix estimation, using the estimates of the matrix elements' moduli and eigenvalues. The main problem is the need to use the optimization criterion, which is different from the maximum likelihood while maximizing the likelihood of the derived matrix. Therefore, the selection of the most appropriate option is performed in Sec. V, where we introduce the results of the Monte Carlo simulations.

### *3.4. Linear Programming Conversion of the Non-Positive Symmetric Toeplitz Matrix into a Positive Definite One with the Specified Minimum Eigenvalue*

Introduced in the previous section, the approach for the estimation of the p.d. Toeplitz symmetric matrix using the moduli of the elements and eigenvalues of the sample matrix relies heavily upon the conversion of the integer-optimized Toeplitz matrix into a p.d. Toeplitz matrix with the specified minimum eigenvalues. Depending on the described option, this transformation ultimately may follow every tested sign change within the covariance matrix diagonals. Note that the eigenvalues pre-processing of the sample matrix included estimation of the number of equal minimal eigenvalues, and this number was then used for the RMT-modification of these eigenvalues. Integer optimization leaves the minimum eigenvalues negative in most cases. Therefore, the final formation of the noise subspace of the p.d. symmetric Toeplitz matrix should be implemented by the following linear programming optimization.

We now consider two options. In the first option, we require the minimum eigenvalue to become positive and equal to the prescribed value. This value may be the a priori known power of the additive white noise or minimum eigenvalue produced by applying the AIC/MDL criteria to the sample matrix $\widehat{\mathbf{R}}_N$, followed by RMT modification. Let us start with the first (simple) LP problem. The main equation used in both LP routines is:

$$\mathbf{T}_N^{(n+1)} = \mathbf{T}_N^{(n)} + \sum_{k=1}^{N-1} \mathbf{A}_k \mathbf{t}_k^{(n+1)} \tag{19}$$

where $\mathbf{A}_k$ is the matrix with only one pair of symmetric diagonals equal to the same element over these two diagonals:

$$\mathbf{A}_k = k \begin{bmatrix} 0 & \dots & 0 & 1 & 0 & \dots & 0 \\ \vdots & & & & 1 & & \\ 0 & & & & & & \\ 1 & & & & & \ddots & \\ 0 & 1 & & & & & 1 \\ & & \ddots & & & & \\ 0 & \dots & 0 & 1 & 0 & \dots & 0 \end{bmatrix} k. \tag{20}$$

For linear programming to be applied, the iterative representation of the eigenvalues of the updated matrix $\mathbf{T}_N^{(n+1)}$ should also be linear, meaning the first-order eigenvalue expansion should be sufficiently accurate. That is possible only for a very small innovation $\mathbf{t}_k^{(n+1)}$ at each step of the iterative representation:

$$-\xi_n < \mathbf{t}_k^{(n+1)} < \xi_n, \tag{21}$$

that keeps the first-order eigenvalues expansion sufficiently accurate:

$$\lambda_j^{(n+1)} = \lambda_j^{(n)} + \sum_{k=1}^{N-1} \mathbf{U}_j^{(n)\mathrm{H}} \mathbf{A}_k \mathbf{U}_j^{(n)} \mathbf{x}_k^{(n+1)}, \tag{22}$$

where $\mathbf{U}_j^{(n)}$ is the $j$-th eigenvector of the matrix $\mathbf{T}_N^{(n)}$. Keeping the first-order eigenvalue expansion accurate is critically important, and therefore, after each $n$-th step of the iterative matrix estimation, the matrix has to be reconstructed using (19), the standard routine for eigenvalues calculations applied with the result compared to the LP solution. Otherwise, the $\xi_n$ in (21) should be reduced, and the LP solution recalculated until it coincides with the MATLAB eigenvalue calculation.

To convert the problem to LP, note that (22) may be presented in a matrix form:

$$\mathbf{\Lambda}_N^{(n+1)} = \mathbf{\Lambda}_N^{(n)} + \mathbf{A}_{N,N-1}^{(n)} \mathbf{X}_{N-1}^{(n+1)}, \tag{23}$$

where

$$\mathbf{A}_{N,N-1}^{(n)} = \begin{bmatrix} \mathbf{U}_1^{(n)\mathrm{H}} \mathbf{A}_1 \mathbf{U}_1^{(n)} & \cdots & \mathbf{U}_1^{(n)\mathrm{H}} \mathbf{A}_{N-1} \mathbf{U}_1^{(n)} \\ \vdots & \ddots & \vdots \\ \mathbf{U}_N^{(n)\mathrm{H}} \mathbf{A}_1 \mathbf{U}_N^{(n)} & \cdots & \mathbf{U}_N^{(n)\mathrm{H}} \mathbf{A}_{N-1} \mathbf{U}_N^{(n)} \end{bmatrix}. \tag{24}$$

For the first problem with the single minimum eigenvalue controlled, let us introduce the $N$-variate vector $\mathbf{X}_N^{(n)}$:

$$\mathbf{X}_N^{(n)} = \begin{bmatrix} \mathbf{X}_{N-1}^{(n)} \\ \cdots \\ y \end{bmatrix}, \tag{25}$$

so that our first LP problem may be formulated as follows:

$$\text{Find min } [0,\dots,0\,|1]\, \mathbf{X}_N^{(n+1)}, \tag{26}$$

subject to:

$$\mathbf{\Lambda}_N^{(n+1)} = \mathbf{\Lambda}_N^{(n)} + \begin{bmatrix} \mathbf{A}_{N,N-1}^{(n)} & \vdots & 1 \\ & \vdots & \vdots \\ & \vdots & 1 \end{bmatrix} \mathbf{X}_{N-1}^{(n+1)}, \tag{27}$$

$$-\boldsymbol{\xi}_{\max} \mathbf{1}_{N-1} < \begin{bmatrix} \mathbf{I}_{N-1} & \vdots & 0 \\ & \vdots & \vdots \\ & \vdots & 0 \end{bmatrix} \mathbf{X}_N^{(n)} < \boldsymbol{\xi}_{\max} \mathbf{1}_{N-1}. \tag{28}$$

The derived LP problem and the need to check the validity of the first-order eigenvalues decomposition represent a considerable amount of calculations required for each LP problem. Yet, this method delivers a single minimal eigenvalue, while the AIC/MDL processing established several noise eigenvalues present. For this reason, to cover all the required and established properties of the optimum solution, let us introduce a more sophisticated LP routine capable of producing the estimated number of noise subspace eigenvalues. Let us subdivide the set of $N$ eigenvalues $\mathbf{\Lambda}_N^{(n)}$ on the three following subsets:

$\mathbf{\Lambda}_{N-K}^{(n)}$ - consists of the different ($N$ – $K$) signal subspace eigenvalues
$\mathbf{\Lambda}_{K-1,1}^{(n)}$ - consists of ($K$ - 1) noise subspace eigenvalues, following the noise subspace
$\mathbf{\Lambda}_{K-1,2}^{(n)}$ - consists of the last ($K$ - 1) noise subspace eigenvalues.

The condition on the equality of the last $K$ minimal eigenvalues is replaced with the equation

$$\mathbf{\Lambda}_{K-1,1}^{(n)} = \mathbf{\Lambda}_{K-1,2}^{(n)}. \tag{29}$$

Let us introduce three matrices:

$\mathbf{A}_{N-K,N-1}^{(n)}$ - consists of the first ($N$ - K) rows of the matrix $\mathbf{A}_{N,N-1}^{(n)}$

$\mathbf{A}^{(n)}_{K-1,N-1,1}$ - consists of the next ($K$ - 1) rows of the matrix $\mathbf{A}^{(n)}_{N,N-1}$

$\mathbf{A}^{(n)}_{K-1,N-1,2}$ - consists of the last ($K$ - 1) rows of the matrix $\mathbf{A}^{(n)}_{N,N-1}$

Then, for each of the three groups of eigenvalues, we have the following equations:

$$\mathbf{\Lambda}^{(n+1)}_{N-K} = \mathbf{\Lambda}^{(n)}_{N-K} + \begin{bmatrix} \mathbf{A}^{(n)}_{N-K,N-1} & \vdots & 1 \\ & \vdots & \vdots \\ & \vdots & 1 \end{bmatrix} \begin{bmatrix} \mathbf{X}^{(n+1)}_{N-1} \\ \cdots \\ y \end{bmatrix} > \lambda_{\min}\mathbf{E}_{N-K}. \tag{30}$$

Since the MATLAB LP routine requires “negative” inequality, after multiplication of (30) by -1, we get:

$$\lambda_{\min}\mathbf{E}_{N-K} - \begin{bmatrix} \mathbf{A}^{(n)}_{N-K,N-1} & \vdots & 1 \\ & \vdots & \vdots \\ & \vdots & 1 \end{bmatrix} \begin{bmatrix} \mathbf{X}^{(n+1)}_{N-1} \\ \cdots \\ y \end{bmatrix} < \mathbf{\Lambda}^{(n)}_{N-K}. \tag{31}$$

Our second equation, which stems from (29), could be written as:

$$\mathbf{\Lambda}^{(n)}_{K-1,1} + \left[\mathbf{A}^{(n)}_{K-1,N-1,1}\right]\mathbf{X}^{(n+1)}_{N-1} - \mathbf{\Lambda}^{(n)}_{K-1,2} - \left[\mathbf{A}^{(n)}_{K-1,N-1,2}\right]\mathbf{X}^{(n+1)}_{N-1} + \mathbf{Z}_k = 0 \tag{32}$$

$$\mu_{\min}\mathbf{1}_{K-1} < \mathbf{Z}_k < \mu_{\max}\mathbf{1}_{K-1}, \tag{33}$$

with the LP optimization that should minimize the following:

$$\text{Find}\min\ (\mu_{\max} - \mu_{\min}). \tag{34}$$

To bring the problem to the canonical (MATLAB) form, let us introduce the ($N$ + $K$ - 1)-variate vector of variables $\mathbf{X}^{(n)}_{N+K-1}$:

$$\mathbf{X}^{(n)}_{N+K-1} = \begin{bmatrix} \mathbf{X}^{(n)}_{N-1} \\ y^{(n)} \\ \mathbf{Z}_{K-1} \\ \mu_{\max} \\ \mu_{\min} \end{bmatrix}. \tag{35}$$

Then, the LP may be formulated as follows:

$$\begin{bmatrix} -\mathbf{A}^{(n)}_{N-K} & \begin{bmatrix} -1 \\ \vdots \\ -1 \end{bmatrix} & \begin{bmatrix} 0 & & \\ & \ddots & \\ & & 0 \end{bmatrix} & \begin{bmatrix} 0 \\ \vdots \\ 0 \end{bmatrix} & \begin{bmatrix} 0 \\ \vdots \\ 0 \end{bmatrix} \\ 0 & 0 & -\mathbf{I}_{K-1} & \begin{bmatrix} 0 \\ \vdots \\ 0 \end{bmatrix} & \begin{bmatrix} 1 \\ \vdots \\ 1 \end{bmatrix} \\ 0 & 0 & \mathbf{I}_{K-1} & \begin{bmatrix} -1 \\ \vdots \\ -1 \end{bmatrix} & \begin{bmatrix} 0 \\ \vdots \\ 0 \end{bmatrix} \\ 0 & 0 & \mathbf{I}_{K-1} & \begin{bmatrix} 0 \\ \vdots \\ 0 \end{bmatrix} & \begin{bmatrix} -1 \\ \vdots \\ -1 \end{bmatrix} \end{bmatrix} \begin{bmatrix} \mathbf{X}^{(n)}_{N-1} \\ y^{(n)} \\ \mathbf{Z}_{K-1} \\ \mu_{\max} \\ \mu_{\min} \end{bmatrix} < \begin{bmatrix} \mathbf{\Lambda}^{(n)}_{N-K} - \lambda_{\min}\mathbf{I}_{N-K} \\ 0 \\ \vdots \\ 0 \end{bmatrix}. \tag{36}$$

Equations with the equality condition are:

$$\begin{bmatrix} \left(\mathbf{A}^{(n)}_{K-1,N-1,1} - \mathbf{A}^{(n)}_{K-1,N-1,2}\right) & \begin{matrix} 0 \\ \vdots \\ 0 \end{matrix} & \mathbf{I}_{K-1} & \begin{matrix} 0 \\ \vdots \\ 0 \end{matrix} & \begin{matrix} 0 \\ \vdots \\ 0 \end{matrix} \end{bmatrix} \times \begin{bmatrix} \mathbf{X}^{(n+1)}_{N-1} \\ y^{(n+1)} \\ \mathbf{Z}^{(n+1)}_{K-1} \\ \mu^{(n+1)}_{\max} \\ \mu^{(n+1)}_{\min} \end{bmatrix} = \mathbf{\Lambda}^{(n)}_{K-1,2} - \mathbf{\Lambda}^{(n)}_{K-1,1}, \tag{37}$$

with

$$-\boldsymbol{\xi}_{(\mathrm{n}+1)}\mathbf{1}_{N-1} \le \mathbf{X}^{(n+1)}_{N-1} \le \boldsymbol{\xi}_{(\mathrm{n}+1)}\mathbf{1}_{N-1}, \quad \mu_{\min} > 0,\ \mu_{\max} > 0. \tag{38}$$

This more elaborate linear programming routine with the same precise control of the first-order eigenvalues expansion accuracy should provide a solution with $K$ equal minimum eigenvalues. Note that the first-order perturbation expansion of the eigenvalues used is accurate for simple eigenvalues and loses its accuracy for eigenvalues that get too close to each other. More sophisticated expansions could be found in [44], for example. Yet, since we used the more straightforward LP problem for our simulations with a single controlled minimum eigenvalue, the number of trials with a first-order expansion failure was minimal, and these trials were excluded from the presented statistics.

Let us repeat that so far, we introduced techniques for the symmetric p.d. Toeplitz matrix reconstruction that only uses the moduli of the sample matrix elements and matrix eigenvalues inspired by the Chu theorem. In Sec. VI, where we introduce the results of the Monte Carlo simulations, we report on the LR value when the criteria of optimization were different from the LR due to the potential phase “calibration” errors impact on the sample matrix $\widehat{\mathbf{R}}_N$. In practical applications, this routine may be used at the initial step for the “calibration” error estimation. At the same time, after these estimates are removed from the sample matrix, the latter could be used for LR maximization. The remaining errors in estimation “calibration” phases lead to lower LR values compared with the case with no “calibration” errors. Since our primary interest is in exploring the ultimate ML estimation accuracy of the symmetric Toeplitz matrix estimation, in our ongoing search for the ML estimate, we assume the absence of “calibration” errors and use the sample covariance matrix $\widehat{\mathbf{R}}_N$ in our algorithms.

## IV. Numerical Techniques for the Likelihood Ratio Maximization, Using the Entire Sample Matrix $\widehat{\mathbf{R}}_N$

In Sec. III, the symmetric Toeplitz matrix was reconstructed using the moduli of the sample matrix $\widehat{\mathbf{R}}_N$ elements and eigenvalues of this matrix that do not depend on the presence of phase errors. Since the sample matrix $\widehat{\mathbf{R}}_N$ could not be used, we had to use different optimization criteria for the matrix reconstruction, though our prime interest is the maximum likelihood estimation of these matrices. For this reason, we consider the problem of maximum likelihood symmetric Toeplitz matrix estimation with a phase-error-free sample matrix available for the LR calculations. The symmetric Toeplitz matrices derived above will be used to initialize the iterative optimization techniques. However, without phase errors, we could use different techniques that adopt this entire sample matrix to generate the initial solution. Part II of this paper introduces some of these techniques devoted to the ML estimation of Hermitian Toeplitz matrices.

The problem of direct ML maximization is an optimization problem, with the probability of converging to the global extremum dependent on initialization. Though other options are available, we use the reconstructed “moduli and eigenvalues” symmetric Toeplitz matrices for initialization. In this traditional approach, we hope to start from a solution with a high probability that belongs to the limited convex sub-area that contains the ML-optimal solution and true Toeplitz covariance matrix. Recall that for $T \to \infty, \widehat{\mathbf{T}}_{\mathrm{ML}} \to \mathbf{T}_N$. The probability of getting the global extremum should increase if we succeed with such an initialization.

In our Part II paper [45], which focuses on the reconstruction of the Hermitian Toeplitz covariance matrices, we analyze an alternative initialization approach based on our ability to establish, in Monte-Carlo simulations, the global nature of the achieved LR value. Specifically, we consider a solution to be the global ML extremum if the two following conditions are met:

1) The LR value produced by this solution with the given sample matrix $\widehat{\mathbf{R}}_N$ exceeds the LR value produced by the true covariance matrix $\mathbf{T}_N$ for the same sample covariance matrix.

2) The ML solution achieved using the true Toeplitz matrix $\mathbf{T}_N$ for initiation and the same sample matrix $\widehat{\mathbf{R}}_N$ coincides with the delivered solution.

These conditions could be applied during the Monte-Carlo simulations where the true Toeplitz matrix is known. Our Part II paper demonstrates that when the true covariance matrix is not known in practical applications, the global ML maximum may also be “recognized” with a sufficiently high probability.

In this paper, we investigate the properties of the derived solutions and use the known true matrix for the identification of the global ML extremum, with no participation of the true covariance matrix $\mathbf{T}_N$ in the optimization algorithm. Driven to get as close as possible to the optimum solution, we propose performing an iterative improvement of the likelihood ratio for the solutions derived in the last section to initiate the optimization.

In Sec. VI, we investigate to what extent the improved LR allows for an improved probability of getting a global extremum using the fmincon routine in direct Monte-Carlo simulations. Also, in some practical cases, the LR improvement achieved by LP may be sufficient for the corresponding problem solution. Recall that the likelihood function for the complex Gaussian data that has to be maximized is [18]

$$\mathrm{LF}(X_1, \dots, X_N \mid \mathbf{T}_N) = ((\det \mathbf{T}_N)^{\mathrm{T}})^{-1} \times \exp\left(-\mathrm{Tr}\left(T\, \widehat{\mathbf{R}}_N \mathbf{T}_N^{-1}\right)\right). \tag{39}$$

For the unknown power $\sigma^2$ of the Toeplitz covariance matrix $\mathbf{T}_N = \sigma^2 \mathbf{T}_N^o$; $[\mathbf{T}_N^o]_{jj} = 1$, it could be replaced by the maximum likelihood estimate (conditional on $\mathbf{T}_N^o$):

$$\hat{\sigma}_{\mathrm{ML}}{}^2 = \frac{1}{T}\mathrm{Tr}\left( \mathbf{R}_N \left(\mathbf{T}_N^{(o)}\right)^{-1}\right), \tag{40}$$

which leads to the so-called “sphericity” test likelihood ratio:

$$\mathrm{LR}\left(X_1, \dots, X_N | \mathbf{T}_N^{(o)}\right) = \frac{\det\left[ \widehat{\mathbf{R}}_N \left(\mathbf{T}_N^{(o)}\right)^{-1}\right]}{\left[\frac{1}{N}\mathrm{Tr}\left( \widehat{\mathbf{R}}_N \left(\mathbf{T}_N^{(o)}\right)^{-1}\right)\right]^{\mathrm{N}}}. \tag{41}$$

Note that the LR in (41) does not depend on $\sigma^2$ of the Toeplitz matrix. While not identically the same, the maximization of the LR in (41) may be interpreted as the search for the most uniform eigenspectrum of the matrix

$$\widehat{\mathbf{R}}_N{}^{1/2}\left(\mathbf{T}_N^{(o)}\right)^{-1}\widehat{\mathbf{R}}_N{}^{1/2}, \tag{42}$$

since the LR in (41) may be treated as the ratio of the mean geometric to the mean arithmetic of the eigenvalues, rising to the power $N$:

$$\mathrm{LR}\left(X_1,\dots,X_N \mid \mathbf{T}_N^{(o)}\right) = \frac{\prod_{j=1}^{N} \mathrm{eig}_j\left(\widehat{\mathbf{R}}_N{}^{1/2}\left(\mathbf{T}_N^{(o)}\right)^{-1}\widehat{\mathbf{R}}_N{}^{1/2}\right)}{\left[\frac{1}{N}\sum_{j=1}^{N} \mathrm{eig}_j\left(\widehat{\mathbf{R}}_N{}^{1/2}\mathbf{T}_N^{(o)}\widehat{\mathbf{R}}_N{}^{1/2}\right)\right]^N}. \quad (43)$$

The LR in (43) reaches its absolute maximum, equal to one, if all eigenvalues are the same. Since this maximum cannot be reached for any finite T in the class of Toeplitz matrices, the "sphericity" test may be interpreted as a specific metric of the difference in eigenvalues. Yet, if we exploit a different criterion of the eigenvalues inequality and try to solve this problem in a small vicinity of the previous solution $\mathbf{T}_N(n)$, we may convert this problem into a sequence of convex optimization problems with a single optimum at each sequence step.

While we should not expect to get the global ML extremum for the LR in (41), we may get closer to it compared with the symmetric matrix we got by integer optimization with the LP modification that used only moduli and eigenvalues of the sample matrix. Moreover, (43) uses the inverted Toeplitz matrix, but since the equalization of the direct matrix eigenvalues leads to some equalization of the inverse eigenvalues, we may try to equalize in small steps the eigenvalues of the matrices $\widehat{\mathbf{R}}_N{}^{-\frac{1}{2}}\mathbf{T}_N(n)\widehat{\mathbf{R}}_N{}^{-\frac{1}{2}}$, inverted with respect to the matrix in (43). To optimize this matrix, we once again should operate with small enough steps $\mathbf{a}_k^{(n)}$ in:

$$\mathbf{T}_N^{(n+1)} = \mathbf{T}_\mathrm{N} + \sum_{k=1}^{N-1} \mathbf{A}_k \mathbf{a}_k^{(n+1)}, \quad -\xi < \mathbf{a}_k < \xi \quad (44)$$

that retain high accuracy of the first-order eigenvalues of the matrix $\mathbf{D} = \widehat{\mathbf{R}}_N{}^{-\frac{1}{2}}\mathbf{T}_N^{(n+1)}\widehat{\mathbf{R}}_N{}^{-\frac{1}{2}}$ expansion:

$$\mathbf{D}_N^{(n+1)} = \widehat{\mathbf{R}}_N{}^{-\frac{1}{2}}\mathbf{T}_N^{(n)}\widehat{\mathbf{R}}_N{}^{-\frac{1}{2}} + \sum_{k=1}^{N-1}\left(\widehat{\mathbf{R}}_N{}^{-1/2}\mathbf{A}_k\widehat{\mathbf{R}}_N{}^{-1/2}\right)\mathbf{a}_k^{(n+1)}, \quad (45)$$

with

$$\lambda_j\left(\mathbf{D}_N^{(n+1)}\right) = \lambda_j\left(\mathbf{D}_N^{(n)}\right) + \sum_{k=1}^{N-1} \mathbf{U}_j^{(n)\mathrm{H}}\left(\widehat{\mathbf{R}}_N{}^{-\frac{1}{2}}\mathbf{A}_k\widehat{\mathbf{R}}_N{}^{-\frac{1}{2}}\right)\mathbf{U}_j^{(n)}\mathbf{a}_k^{(n+1)}, \quad (46)$$

where $\mathbf{U}_j^{(n)}$ is the $j$-th eigenvector of the matrix $\mathbf{D}_N^{(n)}$. Therefore, on the ($n$ + 1)-st iteration, we have to find a small enough $\mathbf{a}_k^{(n+1)}, k = 1,\dots,N-1$, that keeps the first-order eigenvalues decomposition accurate enough by solving the following linear programming problem:

$$\text{Find} \min(\mathbf{X}_{\max} - \mathbf{X}_{\min}) \quad (47)$$

subject to

$$\begin{gathered} \mathbf{X}_{\min}\mathbf{E}_N \le \mathbf{\Lambda}_N^{(n)} + \mathbf{B}_{N,N-1}^{(n)}\mathbf{a}_{N-1}^{(n+1)} \le \mathbf{X}_{\max}\mathbf{E}_N, \\ k = 1,\dots,N, \quad l = 1,\dots,N-1, \\ \mathbf{X}_{\min}, \mathbf{X}_{\max} > 0, \\ -\xi_n \le \mathbf{a}_k^{(n+1)} < \xi_n, \quad \mathbf{E}_N^\mathrm{T} = [1,1,\dots,1]. \end{gathered} \quad (48)$$

It may be formulated as the canonical MATLAB linear programming problem [40]:

$$\text{Find} \min \; [0,\dots,0 \mid -1,+1] \begin{bmatrix} \mathbf{a}_{N-1}^{(n+1)} \\ \cdots \\ \mathbf{X}_{\min} \\ \mathbf{X}_{\max} \end{bmatrix} \quad (49)$$

subject to

$$\begin{bmatrix} -\mathbf{B}_{N,N-1}^{(n)} & \mathbf{E}_N & 0 \\ \mathbf{B}_{N,N-1}^{(n)} & 0 & \mathbf{E}_N \end{bmatrix} \begin{bmatrix} \mathbf{X}_{N-1}^{(n+1)} \\ \cdots \\ \mathbf{X}_{\min} \\ \mathbf{X}_{\max} \end{bmatrix} < \begin{bmatrix} \mathbf{a}_{N-1}^{(n+1)} \\ \cdots \\ \mathbf{X}_{\min} \\ \mathbf{X}_{\max} \end{bmatrix} \quad (50)$$

$$\mathbf{X}_{\min}, \mathbf{X}_{\max} > 0, \quad -\xi_n < \mathbf{a}_k^{(n+1)} < \xi_n. \quad (51)$$

This first-order expansion should be kept accurate by the "flying" constraints $\xi_n$ (51). The check on the validity of the first-order expansion should be conducted at every trial, and the constant $\xi_n$ should be decreased in the LP (49) - (50) and should be recalculated if no suitable match is achieved with the previous $\xi_n$ values. Note that this first-order eigenvalue expansion works for unequal eigenvalues only. With the improved LR proximity of the initial solution to the global LR optimum, we may finally move to the problem of direct LR maximization.

### *4.1. Direct Likelihood Ratio Maximization, Using the MATLAB fmincon Routine*

The optimization techniques introduced above may be treated as producing the initial solutions for the LR maximization routine using MATLAB fmincon optimization. The ultimate goal of the direct LR maximization is to reach the global LR maximum and analyze these solutions' properties. The two criteria introduced above allow for identifying the global LR extremum for the Monte-Carlo simulations with the known true matrix. The properties of the optimum solutions could be applied in practical situations with the unknown true matrix.

The generic MATLAB fmincon routine is introduced in [40] as the non-linear optimization solver that finds the minimum of the problem and is specified as follows [40]:

$$\text{Find}\min f(x) \tag{52}$$

subject to

$$\begin{cases} c(x) \le 0, \\ c_{eq}(x) = 0, \\ \mathbf{A}x < b, \\ \mathbf{A}_{eq}x = b_{eq}, \\ lb < x < ub, \end{cases} \tag{53}$$

where $b$ and $b_{eq}$ are vectors, $\mathbf{A}$ and $\mathbf{A}_{eq}$ are matrices, $c(x)$ and $c_{eq}(x)$ are functions that return vectors, and $f(x)$ is a function that returns a scalar. $f(x)$, $c(x)$ and $c_{eq}(x)$ can be non-linear functions. The problem of direct LR maximization with constraints on the positive definiteness of the optimized matrix fits this software description. Recall that the main reason for the "interim" likelihood maximization was the necessity to have an operational tool for computations that should at least clarify the "benchmark" of many practically important optimizations.

In the following section, we provide a detailed description of the simulation results while exploring the "moduli and eigenvalues" techniques, and we calculate the likelihood ratios for algorithm performance assessment without using LR values for optimization. Since this approach delivers decent solutions capable of successful applications, such as for the phase errors (initial) estimation, in the next section, we use these solutions as the initial ones for the fmincon LR maximization. The "quality" of these initial solutions is defined by the probability of successful trials, where the global maximum likelihood ratio is achieved.

## V. Monte-Carlo Simulations for the "Moduli and Eigenvalues" Methodology of the Symmetric Toeplitz Covariance Matrix Reconstruction

The goal of this chapter is to provide the assessment of the proposed symmetric Toeplitz covariance matrix reconstruction using the "moduli and eigenvalues" of the sample matrix $\widehat{\mathbf{R}}_N$. Apart from the importance of such a reconstruction, for several practical problems, such analysis should validate the Chu theorem for the problems where the input data is represented by a traditional Hermitian sample covariance matrix $\widehat{\mathbf{R}}_N$ averaged over $T$ i.i.d. complex Gaussian random training vectors.

Several general considerations are in order. By exploiting only the "moduli and eigenvalues" of the sample matrix for the reconstruction, we implicitly assume the presence of a beam-steering phase progression and/or phase "calibration" errors that prevent us from using the entire sample matrix for reconstruction. In the presence of these phase errors, the optimum solution to the matrix reconstruction problem cannot reach the ultimate accuracy (in whatever criteria) that can be reached without these interfering factors.

While the maximal LR values that might be achieved in the presence of these interfering factors is an interesting theoretical question, it seems obvious enough that from this reconstruction, one should not expect the same accuracy as in the case of the absence of these interfering factors. Correspondingly, we should not expect to achieve the same (global) maximum for the LR, as per the training data set with no interfering factors.

Another problem is that since the sample covariance matrix cannot be used, apart from the moduli of its elements and eigenvalues not affected by phase errors, the likelihood ratio cannot be calculated. We must use some other "measurable" optimization criteria for the matrix reconstruction. Yet, the connection between the LR and these criteria may be much more complicated than the direct correspondence we would like to observe. Therefore, the non-optimum nature of the optimization algorithm and loose connection with the LR of the actual criteria that may be used for optimization, together with the unknown upper bound on the maximal likelihood ratio that could be achieved, represents both the theoretical and practical problems of the p.d. symmetric Toeplitz reconstruction based on the moduli and eigenvalues of the sample matrix.

For this reason, we introduce the results of the p.d. symmetric matrix reconstruction using our most advanced (and most computationally involved) algorithm, hoping to reveal its potential capabilities. For comparison, we provide the results of this algorithm when the LR is the optimization criterion. If the sample matrix $\widehat{\mathbf{R}}_N$ is available for the LR calculations, then there is no reason to confine this matrix restoration only by the moduli of these matrix elements and their eigenvalues.

Therefore, while the practical value of this technique is close to none, the results of this optimization provide important data analysis for comparison with the "operational" algorithms mentioned above that do not use the entire sample matrix $\widehat{\mathbf{R}}_N$ and therefore, do not use the LR values in the optimization process.

In this algorithm, we start from the "integer dynamic programming algorithm," using the $L_2$ norm of the difference between the eigenvalues specified by RMT and the eigenvalues of the reconstructed matrix as the optimization criterion. Due to the "naïve" moduli estimates used at this stage, a certain number of negative minimum eigenvalues are always present in the solutions with these moduli.

While the original integer-only option was also explored with no significant success, we augmented every trial with the sign inversion distributed over the sub-arrays by the linear programming routine that converts the non-positive definite Toeplitz symmetric matrix into a positive definite one with the specified minimum eigenvalues. Therefore, a comparison of the optimization criterion for different positions of the sign change along the matrix's sub- (and super-) diagonals is performed for a p.d. Toeplitz matrix so that along with the criterion, the LR may also be calculated.

In the first stage of our optimization, following the “integer dynamic programming” from [21] and [22], we form ($N$ - 1) “branches” with the sign change corresponding to this “branch’s” sub- (and super-) diagonals. Then, the tests for sign inversion positions in each branch are accompanied by LP conversion of the tested matrix into a positive definite matrix with the prescribed minimal eigenvalues.

These matrices, implemented with different selected sub-diagonals from the sign inversion, get compared by finding the best position that maximally reduces the $L_2$ norm between the specified eigenvalues and eigenvalues of the optimized matrix. For these p.d. symmetric Toeplitz matrices, we may also calculate the likelihood ratio, which is not a part of optimization but helps to investigate the discrepancy between the maximum LR and minimum $L_2$ eigenvalues distinction criteria. When this initial distribution of sign inversions is performed in each “branch,” we conduct a second optimization within each branch instead of selecting the best branch.

In the second optimization, we reconstruct the oldest sign inversion and look again for the best position for this sign change, keeping all other previous sign changes intact. Each trial is accompanied by the LP conversion of the non-positive definite matrix into a positive definite one with the specified minimum eigenvalue.

Since our first-order eigenvalue expansion (22) is accurate only for all different eigenvalues, in cases when the LP cannot convert a matrix with a few negative eigenvalues into a p.d. matrix, we can apply diagonal loading to make the minimum eigenvalue equal to the prescribed value. This allows us to constantly deal with p.d. Toeplitz matrices and calculate the LR value. We also adopted a new criterion of optimization. In addition to the minimum $L_2$ difference of eigenvalues, we use the minimax criterion

$$\min_{n} \max_{j} \frac{\left|\lambda_j(n) - \lambda_j^*\right|}{\lambda_j^*}, \tag{54}$$

where $\lambda_j^*$ is the RMT-specified $j$-th eigenvalue. We provided a rather detailed description of our algorithm to clarify that it may be modified in several ways. Yet, we expect the algorithm to provide the correct estimate of what could be achieved with the “moduli and eigenvalues” data and optimization criterion, which differs from the likelihood ratio.

Let us now analyze the optimization results for several trials. In our first example (TABLE II. ), out of 17 “branches,” 11 finished with the same $L_2$ eigenvalues distance, and two other results were repeated three times each.

TABLE II.

| L2 Eigenvalues Dist. | LR | minimax |
|---|---|---|
| 22.83 | 0.0002 | 0.1245 |
| 5.78 | 0.0008 | 0.1310 |
| 5.78 | 0.0008 | 0.1230 |
| 5.78 | 0.0008 | 0.1269 |
| 9.09 | 4.56E-07 | 0.1305 |
| 9.09 | 4.56E-07 | 0.1313 |
| 9.09 | 4.56E-07 | 0.1308 |
| 9.40 | 0.0017 | 0.1209 |
| 9.40 | 0.0017 | 0.1238 |
| 9.40 | 0.0017 | 0.1229 |
| 9.40 | 0.0017 | 0.1286 |
| 9.40 | 0.0017 | 0.1284 |
| 9.40 | 0.0017 | 0.1219 |
| 9.40 | 0.0017 | 0.1332 |
| 9.40 | 0.0017 | 0.1277 |

Therefore, the $L_2$ criterion seems insufficient to select the best solution. The three solutions with the minimal $L_2$ value (5.78) all have the same LR values (0.008), which is not the best one, and the minimax eigenvalue criterion is not the best either (0.1230, 0.1310). The maximum LR (0.0017) is achieved in all eight solutions with the same $L_2$ = 9.40, and only the minimum of the minimax criterion (54) provided the maximum LR = 0.0017.

In our second example (TABLE III. ), the minimum of $L_2$ = 0.2130 was repeated seven times, and all of these solutions had a very low LR value (LR = $4.01 \cdot 10^{-5}$).

TABLE III.

| L2 Eigenvalues Dist. | LR | minimax |
|---|---|---|
| 0.2261 | 0.0001 | 0.1310 |
| 0.8067 | 6.66E-05 | 0.1310 |
| 0.2130 | 4.01E-05 | 0.1335 |
| 0.2130 | 4.01E-05 | 0.1387 |
| 0.2130 | 4.01E-05 | 0.1319 |
| 0.3278 | 0.0017 | 0.1277 |
| 0.3278 | 0.0017 | 0.1251 |
| 0.2563 | 0.0000 | 0.1537 |
| 0.2498 | 0.0008 | 0.1255 |
| 0.2261 | 0.0007 | 0.1214 |
| 0.3278 | 0.0017 | 0.1229 |
| 0.2130 | 4.01E-05 | 0.1336 |
| 0.2130 | 4.01E-05 | 0.1316 |
| 0.2130 | 4.01E-05 | 0.1283 |
| 0.2130 | 4.01E-05 | 0.1242 |

The minimum minimax criterion (0.1214) produced a poor LR = 0.0007, compared with the maximum LR = 0.0017. The two best LR solutions are not the best in $L_2$ and minimax criterion. In our third example (TABLE IV. ), the minimax criterion (0.1043) coincided with LR = 0.0124, though this LR value was present in four other solutions that are not optimal in any other criterion.

TABLE IV.

| L2 Eigenvalues Dist. | LR | minimax |
|---|---|---|
| 0.2463 | 3.30E-28 | 0.1240 |
| 0.8055 | 5.34E-10 | 0.1295 |
| 0.2463 | 1.33E-05 | 0.1247 |
| 0.2463 | 1.33E-05 | 0.1248 |
| 0.2463 | 1.33E-05 | 0.1253 |
| 0.4627 | 0.0027 | 0.1342 |
| 0.4627 | 0.0027 | 0.1284 |
| 0.2151 | 0.0124 | 0.1087 |
| 0.2151 | 0.0124 | 0.1135 |
| 0.2151 | 0.0124 | 0.1043 |
| 0.2305 | 0.0154 | 0.1130 |
| 0.2305 | 0.0154 | 0.1126 |
| 0.2305 | 0.0154 | 0.1147 |
| 0.2305 | 0.0154 | 0.1146 |
| 0.2305 | 0.0154 | 0.1138 |

In our fourth example (TABLE V. ), the minimax and max LR solutions are again the same.

TABLE V.

| L2 Eigenvalues Dist. | LR | minimax |
|---|---|---|
| 28.33 | 0.0006 | 0.1197 |
| 5.78 | 0.0048 | 0.1246 |
| 5.78 | 0.0048 | 0.1250 |
| 5.78 | 0.0048 | 0.1239 |
| 5.78 | 0.0048 | 0.1257 |
| 5.78 | 0.0048 | 0.1289 |
| 5.78 | 0.0048 | 0.1216 |
| 5.78 | 0.0048 | 0.1239 |
| 5.78 | 0.0048 | 0.1276 |
| 5.78 | 0.0048 | 0.1249 |
| 5.78 | 0.0048 | 0.1243 |
| 5.78 | 0.0048 | 0.1203 |
| 5.78 | 0.0048 | 0.1297 |
| 5.78 | 0.0048 | 0.1221 |
| 5.78 | 0.0048 | 0.1222 |

These and many similar examples demonstrated that no "working" criterion is entirely identical to the maximum likelihood criterion, and the closest to the ML criterion may be the minimax criterion. For this reason, to demonstrate the best LR values achieved using the "moduli and eigenvalues" of the sample matrix, we selected the final results delivered by the minimax criterion, which was applied in the second stage of our optimization.

To clarify the proposed optimization technique's potential capability, we provide the same process's results with the maximum likelihood criterion used in each procedure step.

While this option does not have a practical value, it demonstrates the optimization limitations and the unrelated likelihood ratio criteria. Comparison with the LR optimization results (Sec. VI) demonstrates the LR losses this "integer-non-integer" algorithm has, compared with the technique that delivers the global LR maximum.

The data in Fig. 1 are averaged over 33 trials only since the integer sign change, followed by LP conversion of the resulting non-positive definite matrix into a p.d. matrix with the specified minimum eigenvalue required for the LR calculation for every sign change, is a very time-consuming procedure.

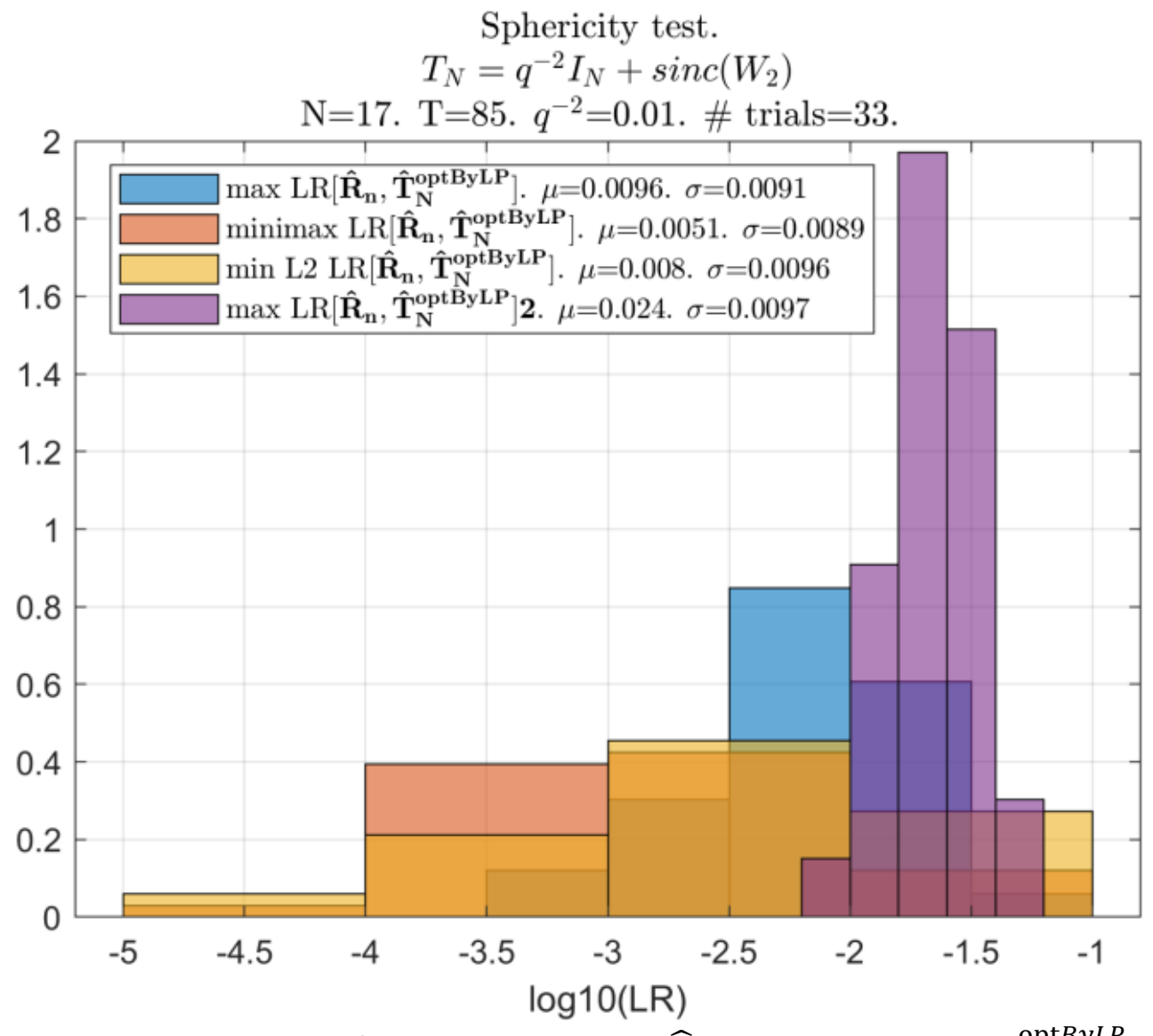

Fig. 1. Sphericity test after 2 optimizations. $\hat{\mathbf{R}}_N$ = sample matrix, $\mathbf{T}_N^{optByLP}$ = covariance matrix after LP optimizations. Max LR = max LR after 1st optimization, minimax LR = LR with min minimax, min CR LR = LR with min-max eigenvalue dist., max LR 2 = max LR after 2nd optimization. $\mathbf{T}_N$ = $q^{-2}\mathbf{I}_N$+ sinc ($W_2$). N=17, T=85, 33 trials.

Still, comparing this LR maximization results with the results of the "practical" symmetric positive definite Toeplitz matrix restoration provides the required data for reliable observations. Specifically, this procedure of LR optimization by the proposed "integer-non-integer" routine provides much better LR values than the compared "working" criteria. On the other hand, these results remain inferior to the "expected likelihood" generated by the true covariance matrix.

For this optimization, we got the LR pdf, confined to the interval 0.01 < LR < 0.1, while the likelihood ratio of the true covariance matrix is within the interval 0.1 < LR < 0.22. Inverse losses may be attributed to the specifics of the "integer-non-integer" optimization. The results of this LR optimization significantly outperform our "practical" algorithms that use different optimization criteria.

All three optimizations shared the same "first-order" optimization, performed by the "integer dynamic programming" algorithm with no linear programming, applied to convert the solutions into positive definite ones. The optimization within each "branch" of dynamic programming was conducted using the minimum of the $L_2$ norm of the eigenvalues discrepancy between the eigenspectrum of the optimized Toeplitz matrix and the RMT-modified eigenvalues of the sample matrix $\widehat{\mathbf{R}}_N$.

All distinctions were introduced for the "second-order" optimization of every "branch" solution of the pure integer optimization. Here, we applied the LP conversion of the non-positive definite Toeplitz matrix into a p.d. one with the specified minimum eigenvalue, and for the three introduced pdfs, we applied different optimization criteria.

The first criterion was the same $L_2$ norm between the eigenvalues of the reconstructed Toeplitz matrix and the RMT-specified ones. The second was the minimax criterion that minimized the maximal over 17 eigenvalue distance

$$\min \max_j \frac{|\lambda_j^* - \lambda_j(n)|}{\lambda_j^*}, \quad (55)$$

where $\lambda_j^*$ is the given set of RMT-modified eigenvalues of the sample matrix, and $\lambda_j(n)$ is the $j$-th eigenvalue of the reconstructed Toeplitz matrix. Our third criterion was based on the eigenspectrum of the matrix $\mathbf{D}(n)$:

$$\mathbf{D}(n) = \frac{1}{T}\sum_{t=1}^{T} \mathrm{diag}(\mathbf{x}_t^*)\mathbf{T}(n)^{-1}\,\mathrm{diag}(\mathbf{x}_t). \quad (56)$$

The eigenspectrum of this matrix is not affected by the phase errors, while in the absence of these errors, we get

$$\mathbf{1}_N^T\mathbf{D}(n)\mathbf{1}_N = \frac{1}{T}\sum_{t=1}^{T} \hat{\sigma}_t^2\,; \quad \hat{\sigma}_t^2 = \mathbf{x}(t)^{\mathrm{H}}\mathbf{T}(n)^{-1}\mathbf{x}(t). \quad (57)$$

For this reason, we introduced the following criterion for the selection:

$$\rho = \frac{\lambda_{\max}[\mathbf{D}(n)] - \lambda_{\min}[\mathbf{D}(n)]}{\sum_{j=1}^{N}\lambda_j[\mathbf{D}(n)]}. \quad (58)$$

The results of these 33 trials allowed us to compare the efficiency of the proposed algorithms for the (impractical) LR maximization that produced the mean LR level $\mu = 0.024$ with these three "working" criteria:

$$\begin{gathered}\min L_2 \text{ norm - } \mu = 0.0051,\\ \min \rho \text{ - } \mu = 0.0091,\\ \text{min-max - } \mu = 0.0096.\end{gathered} \quad (59)$$

This comparison demonstrates the superiority of the (impractical) LR maximization ($\mu = 0.024$) over all tested "practical" criteria, with the best result being (55). The best of the three "practical" criteria can be selected based on the results of the simulations (shown below) with 1000 trials for each tested sample volume $T$. The pdf generated by the true Toeplitz covariance matrix $\mathbf{T}_N$ ("expected likelihood") is also introduced in Fig. 2 for comparison.

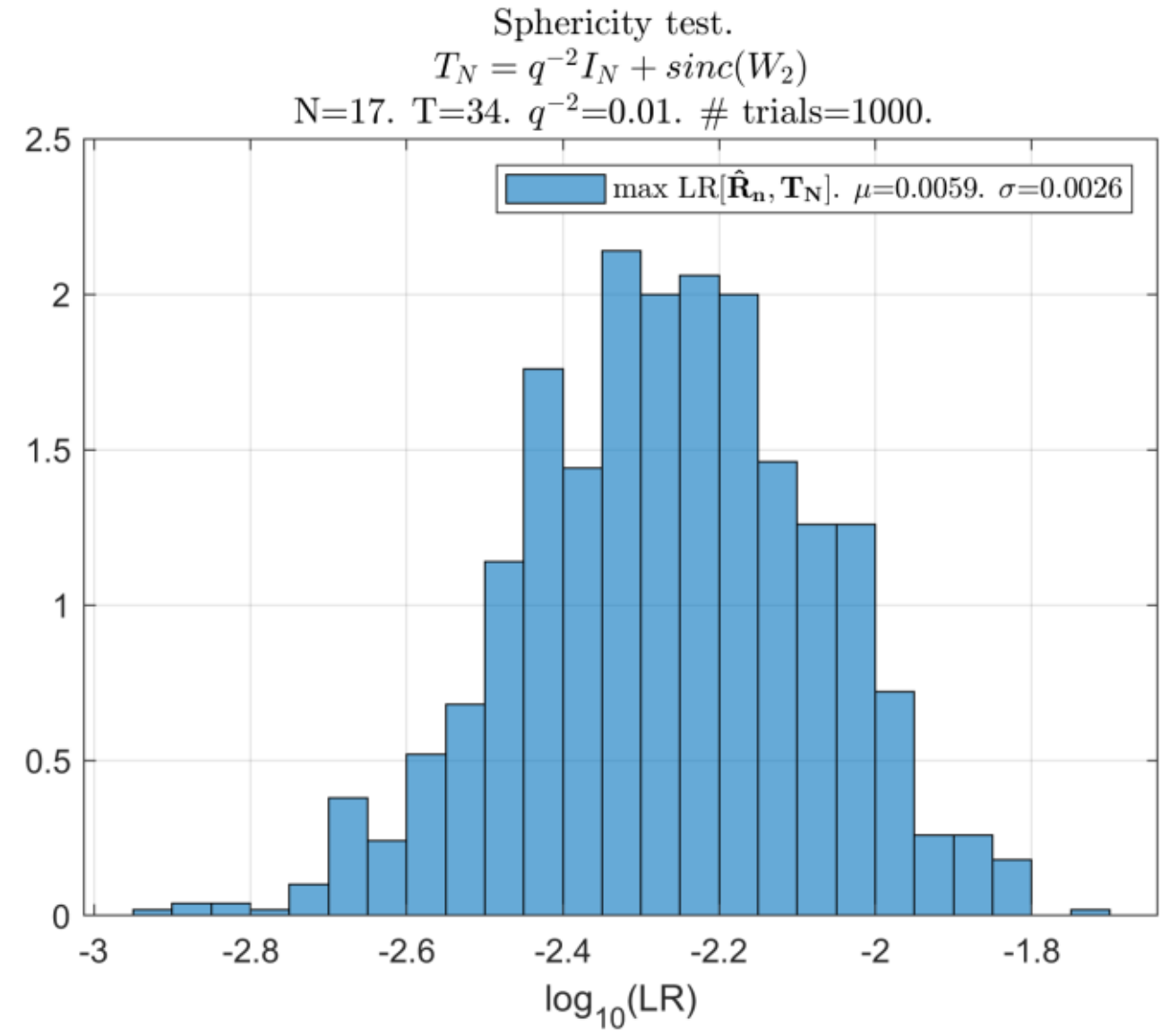

Fig. 2. Sphericity test for true Toeplitz covariance matrix. $\widehat{\mathbf{R}}_\mathbf{N}$ = sample matrix, $\mathbf{T}_\mathbf{N}$ = covariance matrix. $\mathbf{T}_\mathbf{N} = q^{-2}\mathbf{I}_\mathbf{N} + \mathrm{sinc}(W_2)$. N=17, T=34, 1000 trials.

In Fig. 3 - Fig. 7, we provide the sample pdfs of the likelihood ratio for the optimization criteria used in the second stage of our optimization.

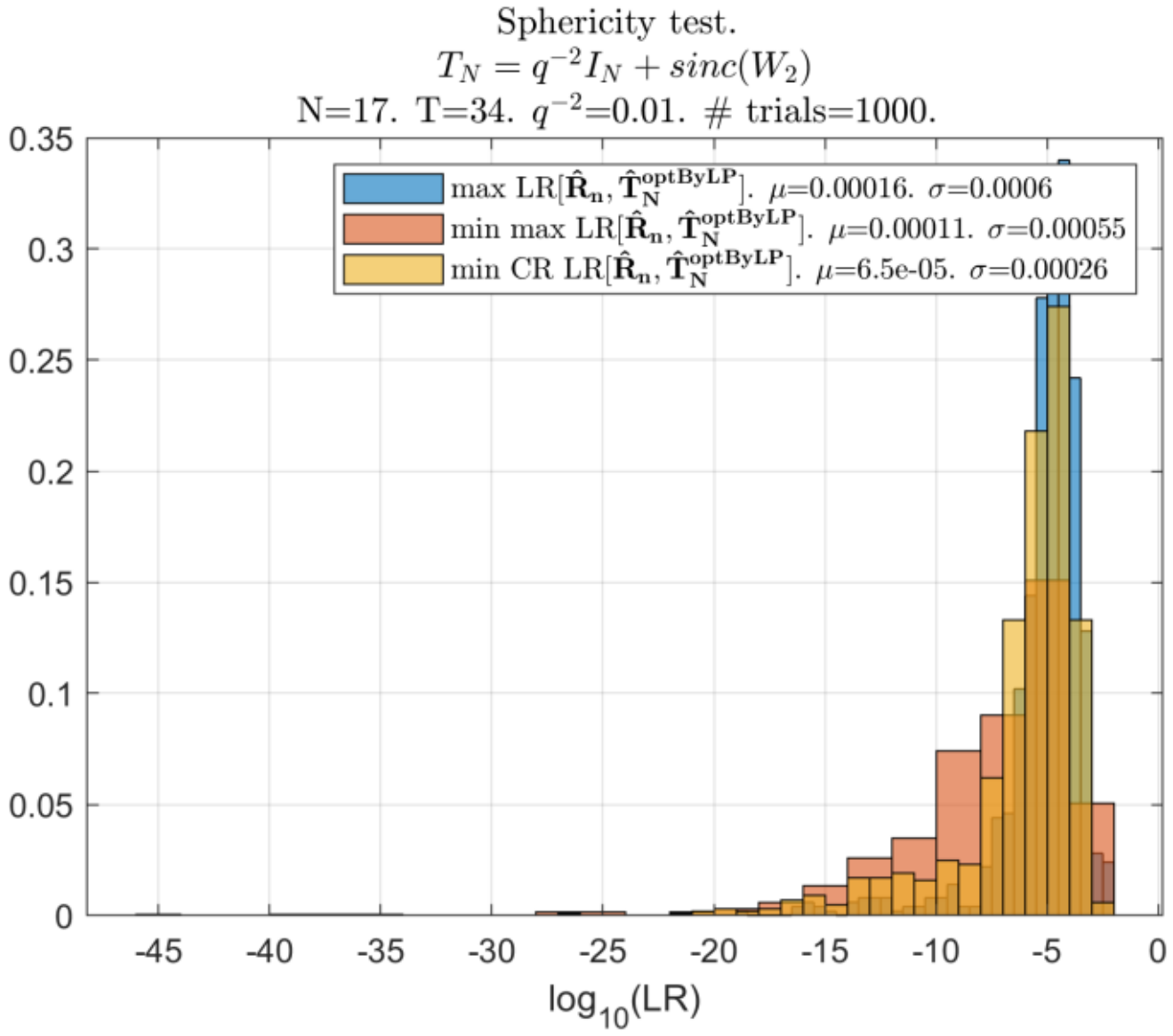

Fig. 3. Sphericity test after 2 optimizations. $\widehat{\mathbf{R}}_\mathbf{N}$ = sample matrix, $\mathbf{T}_\mathrm{N}^{\mathrm{opt}ByLP}$ = covariance matrix after LP optimizations. Max LR = max LR after 1st optimization, minimax LR = LR with min minimax, min CR LR = LR with min-max eigenvalue dist. $\mathbf{T}_\mathbf{N} = \mathrm{q}^{-2}\mathbf{I}_\mathrm{N} + \mathrm{sinc}\,(\mathrm{W}_2)$. N=17, T=85, 1000 trials.

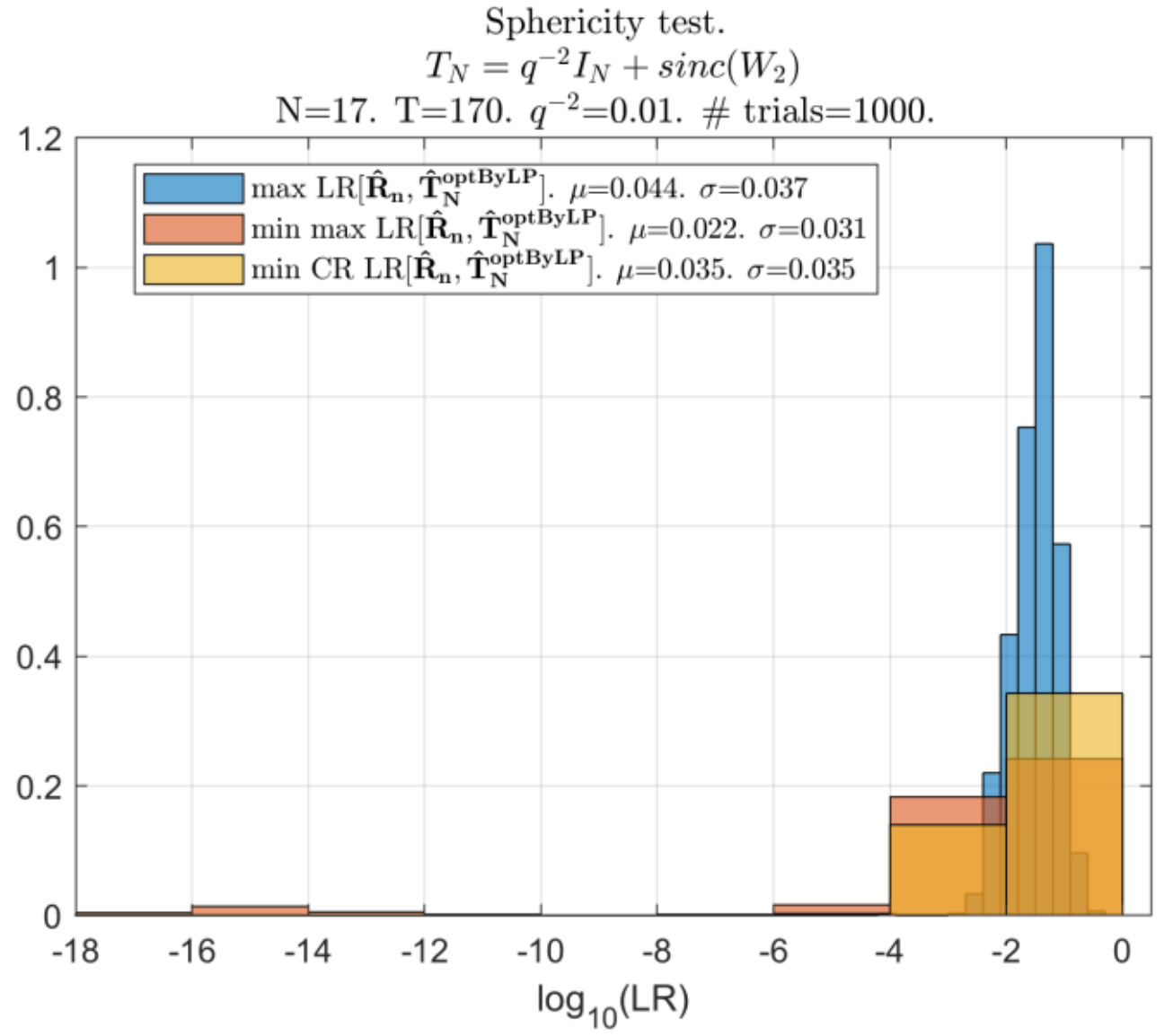

Fig. 5. Sphericity test after 2 optimizations. $\widehat{\mathbf{R}}_\mathbf{N}$ = sample matrix, $\mathbf{T}_\mathrm{N}^{\mathrm{opt}ByLP}$ = covariance matrix after LP optimizations. Max LR = max LR after 1st optimization, minimax LR = LR with min minimax, min CR LR = LR with min-max eigenvalue dist. $\mathbf{T}_\mathbf{N} = \mathrm{q}^{-2}\mathbf{I}_\mathrm{N} + \mathrm{sinc}\,(\mathrm{W}_2)$. N=17, T=170, 1000 trials.

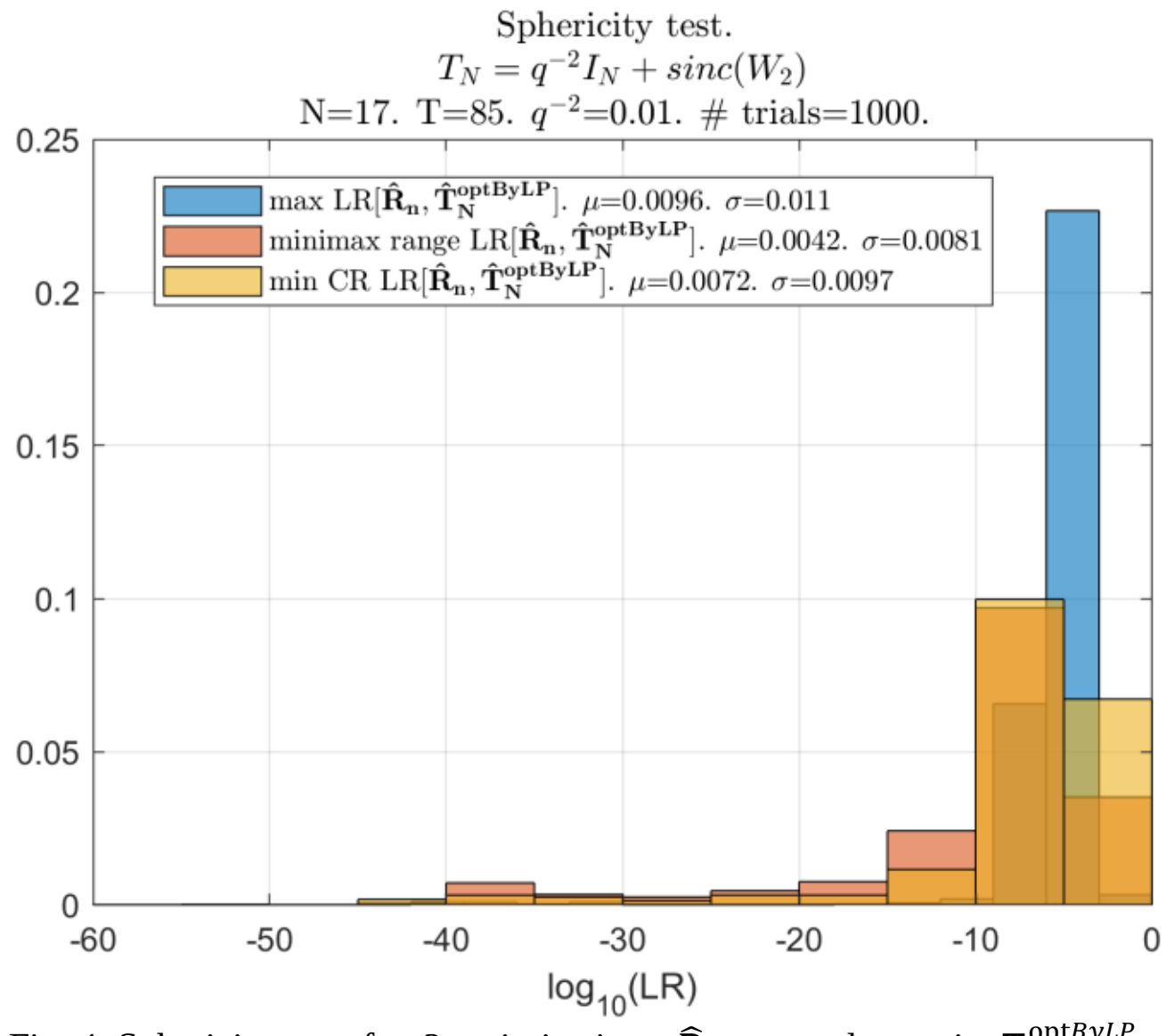

Fig. 4. Sphericity test after 2 optimizations. $\widehat{\mathbf{R}}_\mathbf{N}$ = sample matrix, $\mathbf{T}_\mathrm{N}^{\mathrm{opt}ByLP}$ = covariance matrix after LP optimizations. Max LR = max LR after 1st optimization, minimax LR = LR with min minimax, min CR LR = LR with min-max eigenvalue dist. $\mathbf{T}_\mathbf{N} = \mathrm{q}^{-2}\mathbf{I}_\mathrm{N} + \mathrm{sinc}\,(\mathrm{W}_2)$. N=17, T=85, 1000 trials.

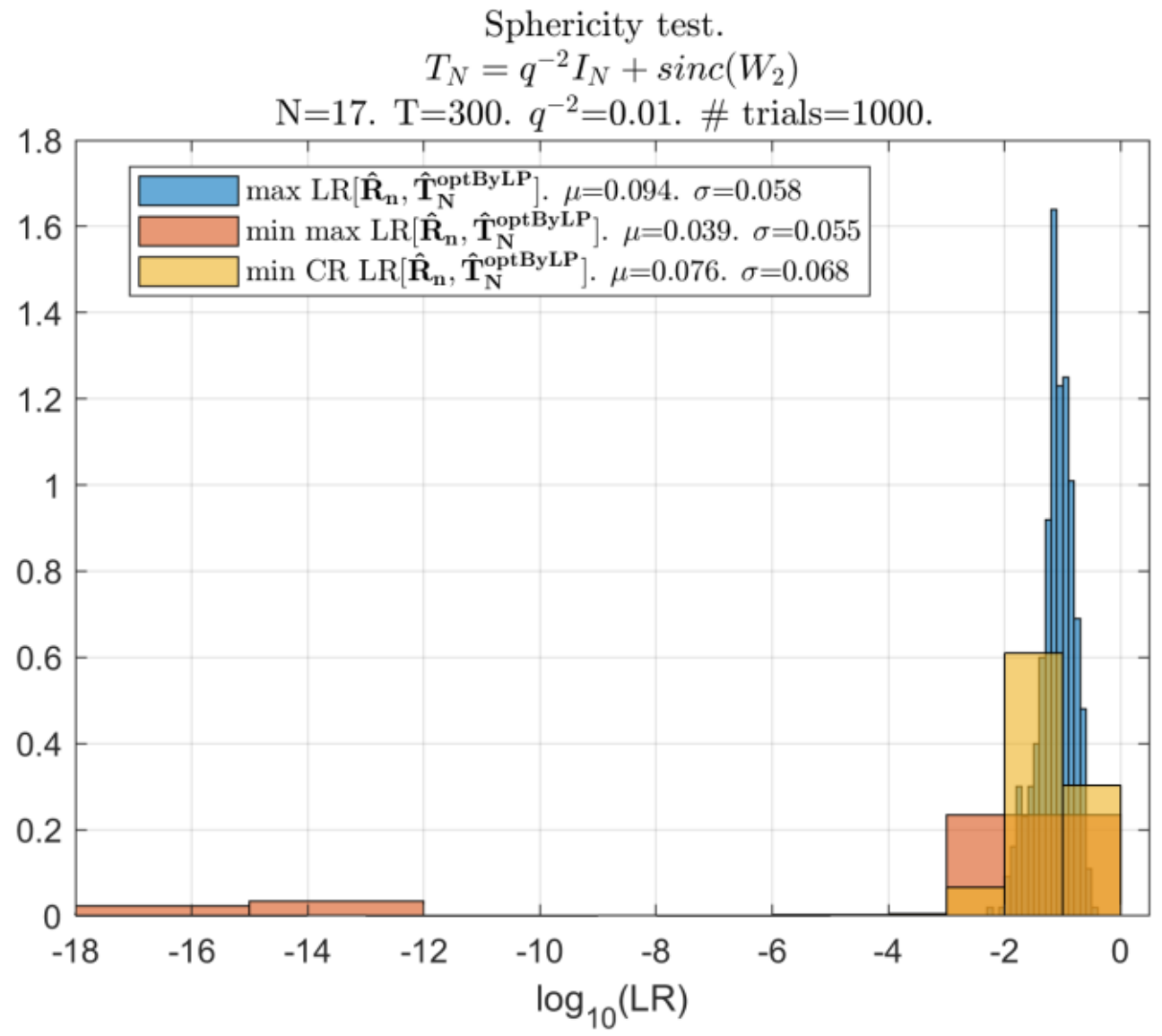

Fig. 6. Sphericity test after 2 optimizations. $\widehat{\mathbf{R}}_\mathbf{N}$ = sample matrix, $\mathbf{T}_\mathrm{N}^{\mathrm{opt}ByLP}$ = covariance matrix after LP optimizations. Max LR = max LR after 1st optimization, minimax LR = LR with min minimax, min CR LR = LR with min-max eigenvalue dist. $\mathbf{T}_\mathbf{N} = \mathrm{q}^{-2}\mathbf{I}_\mathrm{N} + \mathrm{sinc}\,(\mathrm{W}_2)$. N=17, T=300, 1000 trials.

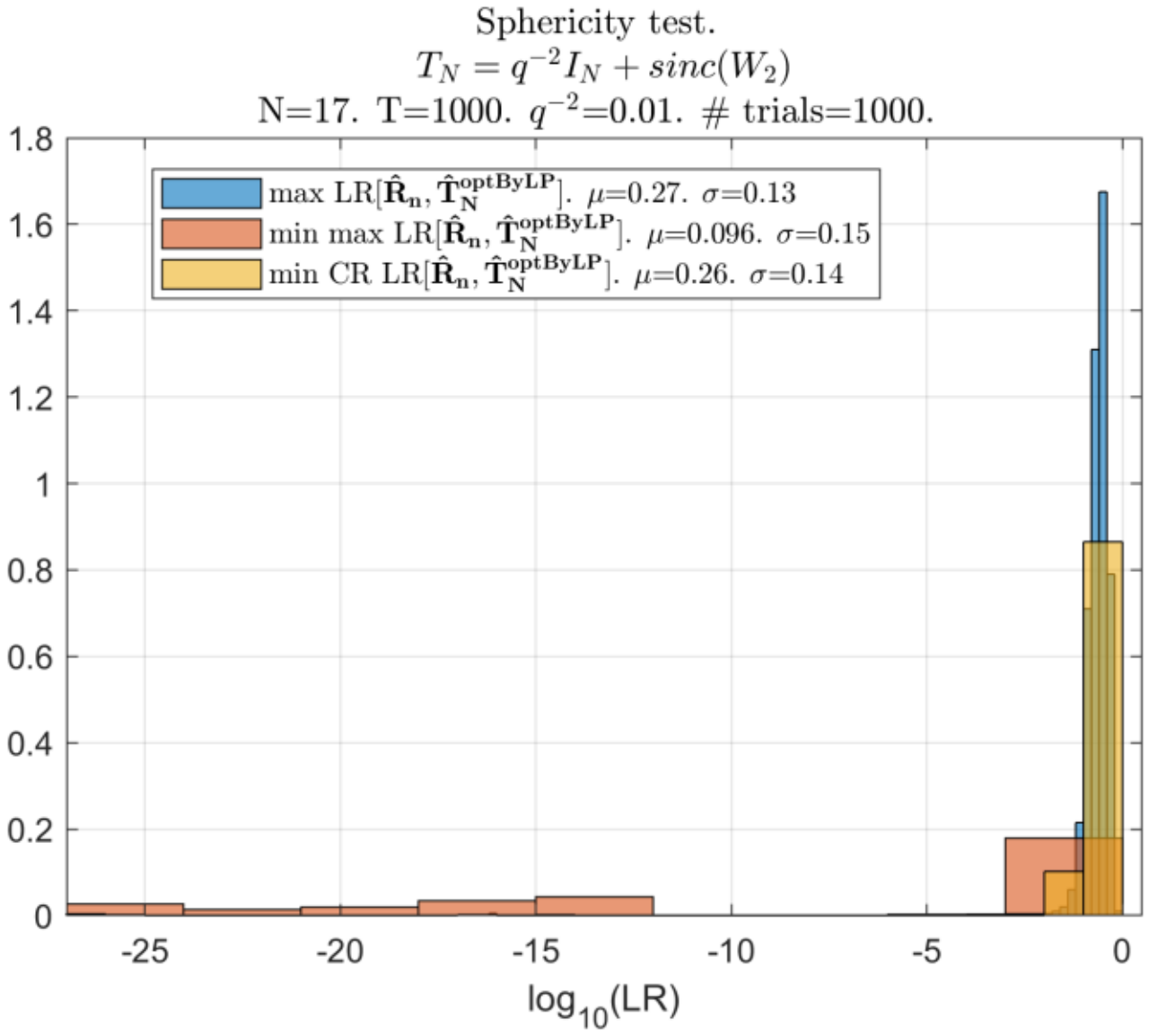

Fig. 7. Sphericity test after 2 optimizations. $\widehat{\mathbf{R}}_\mathbf{N}$ = sample matrix, $\mathbf{T}_\mathrm{N}^{\mathrm{opt}ByLP}$ = covariance matrix after LP optimizations. Max LR = max LR after 1st optimization, minimax LR = LR with min minimax, min CR LR = LR with min-max eigenvalue dist. $\mathbf{T}_\mathrm{N} = \mathrm{q}^{-2}\mathbf{I}_\mathrm{N} + \mathrm{sinc}\,(\mathrm{W}_2)$. N=17, T=1000, 1000 trials.

The first stage of the "integer dynamic programming" was the same for all these options, and it was implemented by pure "integer dynamic programming" with the remaining negative minimal eigenvalues. For this reason, when the LR criterion is applied in the second optimization stage, the results are worse than for the properly optimized distributed sign changes of the first stage (Fig. 1).

Still, they are markedly better than the LR values optimized using the LR criterion with the LP matrix conditioning at both stages of optimization. Also, we have to admit that the results of 1000 trials did not support our expectations regarding the minimax optimization criterion: the standard min $L_2$ norm of all eigenvalues discrepancy does not seem to be worse. Yet, these results confirm the losses in LR when other optimization criteria were used.

The Toeplitz matrix reconstruction using the "moduli and eigenvalues" of the sample matrix $\widehat{\mathbf{R}}_N$ provides the p.d. symmetric Toeplitz matrices with LR up to the order of magnitude (for the median sample volumes $T$) lower than the "expected likelihood" of the true covariance matrix $\mathbf{T}_N$. We should not be surprised that the "moduli and eigenvalues" symmetric Toeplitz matrix restoration, using the "LR-related" optimization criteria, provided comparatively low LR values.

For a good number of applications, the reconstructed Toeplitz matrices are accurate enough to provide the required efficiency of the problem solution. In particular, these estimates could be used for the initial "calibration" phase errors estimation, followed by re-estimation of the symmetric Toeplitz matrix with the corrected sample matrix $\widehat{\mathbf{R}}_N$, now available for optimization. The LR losses we observe are caused by applied "practical" optimization criteria, different from LR, and by the limitations associated with only two (mostly incompatible) sets of estimates provided. It should be kept in mind that the maximal LR, equal to one, is generated by the sample matrix $\widehat{\mathbf{R}}_N$, while the true covariance matrix $\mathbf{T}_N$ generates an "expected likelihood" much smaller, especially for a limited testing sample volume.

## VI. Monte-Carlo Simulations Results on LR Maximization, Using the Phase Errors-Free Sample Covariance Matrix

In this section, we introduce the results of the LR maximization for the symmetric Toeplitz covariance matrix when the "calibration" phase errors are known to be absent. Since our "moduli and eigenvalues" Toeplitz matrix reconstruction methodology cannot deliver the globally optimum (maximum likelihood) results, we considered performing the continuing LR maximization in two steps. In the first step, we planned to use the LP routine that should tend to equalize the eigenvalues of the matrix $\widehat{\mathbf{R}}_N^{-\frac{1}{2}}\mathbf{T}(n)\widehat{\mathbf{R}}_N^{-\frac{1}{2}}$, in expectation that such equalization should enhance the LR of the symmetric Toeplitz matrix, increasing the chances of getting the globally optimum solution by the LR optimization using the MATLAB fmincon routine. In reality, we discovered that by using the results of the "moduli and eigenvalues" optimization as the initial solutions, the fmincon routine always converged to the globally optimal ML solution, apart from the rare cases when it converged to non-positive definite Toeplitz matrices.

In what follows, we report on the probability of the fmincon failure as a function of the training sample volume T. Still, the need to decrease the probability of convergence to a (legitimate) local LR extremum did not exist throughout all conducted simulations.

For this reason, we do not introduce the results of the Monte-Carlo simulations for this algorithm while leaving its description for its possible use as an alternative to the fmincon routine. Our general observation is that LR improvement, delivered by this LP optimization, relied upon the first-order eigenvalue expansion, which was relatively modest.

### *6.1 Monte-Carlo Simulation results on the Likelihood Ratio Maximization, Using the MATLAB fmincon Routine*

Let us analyze the results of the MATLAB fmincon likelihood ratio maximization using the solutions of the "moduli and eigenvalues" LR optimization as the initial solutions for the fmincon optimization.

The simulations have been conducted for the sample volumes $T$ = 17, 34, 85, 170, 240, 300, 340, 510, 850, 1000, 2000, 3000, 5000, 7000, and 10000, with 1000 Monte-Carlo trials for each sample volume. As mentioned above, in some cases, fmincon converged to a non-positive definite Toeplitz matrix. In Fig. 8, we provide the percentage of such events.

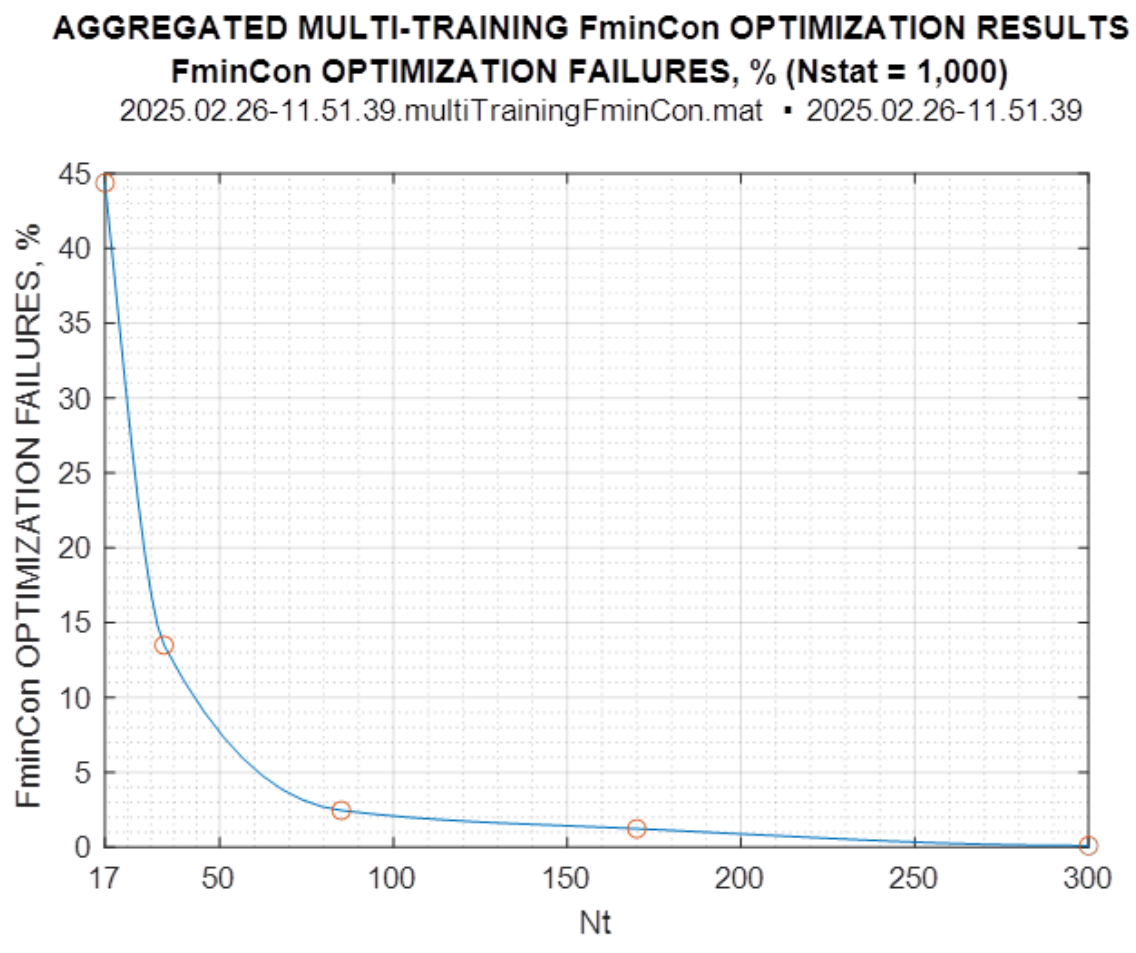

Fig. 8. Fmincon optimization failures, %

A maximum of 45% of all 1000 trials was observed for the minimal sample volume $T = 17$, but already for $T = 85$, there were less than 2% of such trials. The statistics below are provided for 1000 "successful" trials, all meeting our two criteria for being the global LR maximum.

Let us first illustrate the convergence of the fmincon optimization. In Fig. 9, we present the LR values and minimum (noise) eigenvalue as a function of the iteration number.

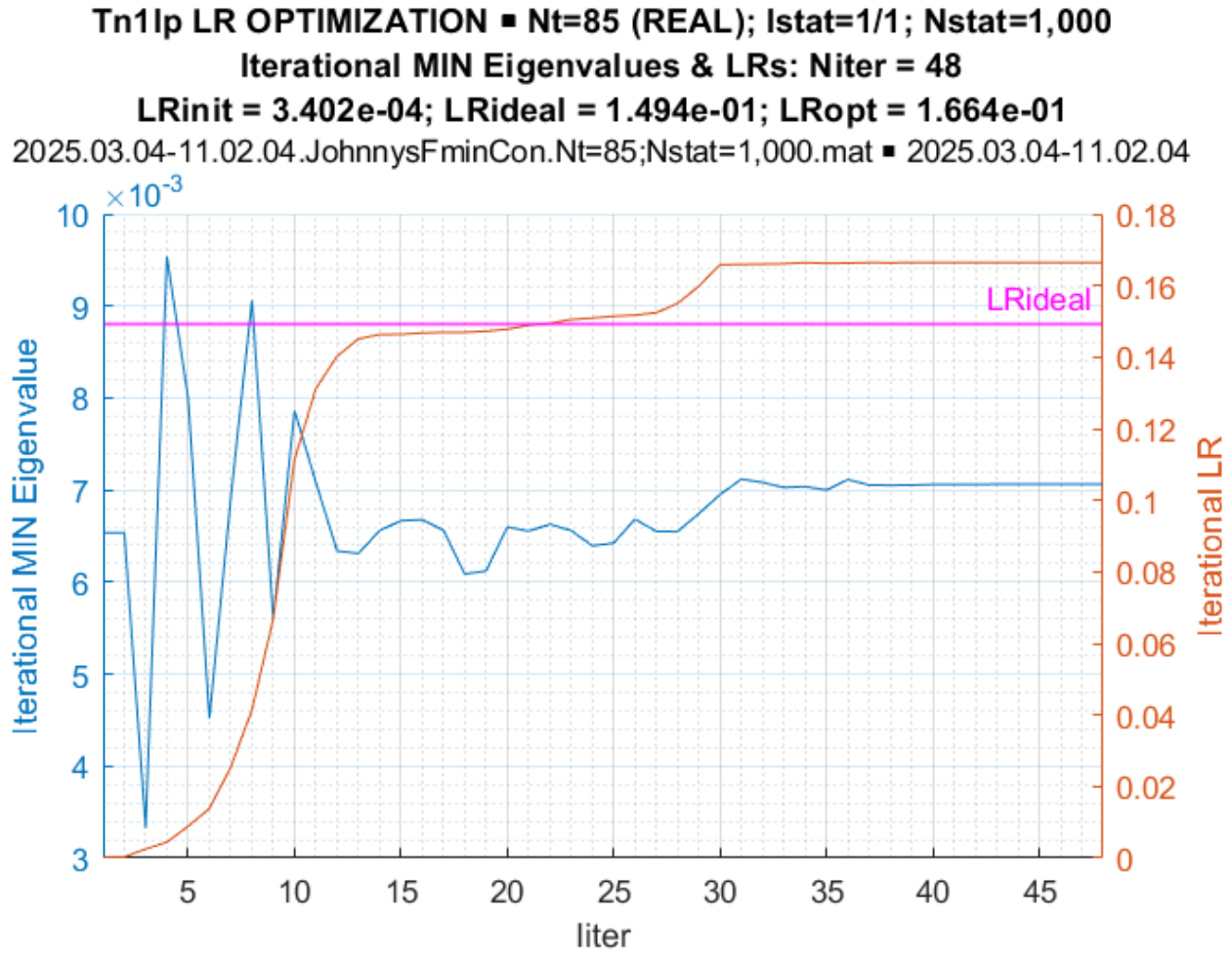

Fig. 9. Iterative LR and Minimum eigenvalues while LR optimization

Here, the "expected likelihood" of the true covariance matrix is presented as the line going through LR = 0.15. One can see that at the 15th iteration, the procedure reached the "expected likelihood" level, and by the 30th iteration, it converged to its final value, both in terms of the likelihood ratio LR = 0.17 > 0.15 and the minimum eigenvalue ($\lambda_{\min} = 7 \cdot 10^{-3}$). Fig. 10 illustrates the convergence of the (normalized) eigenspectrum of the optimized Toeplitz matrix.

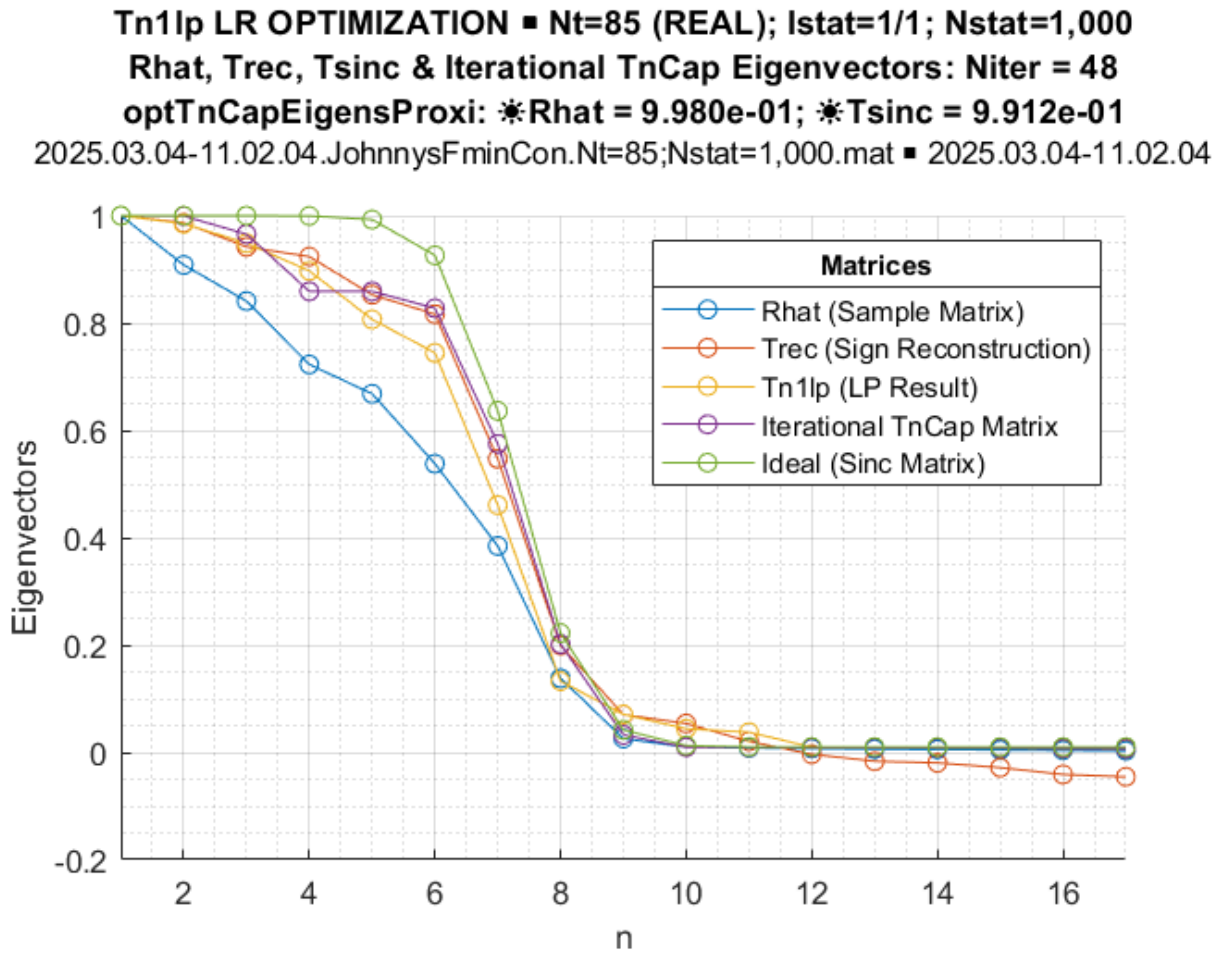

Fig. 10. Processed matrices normalized eigenvalues

The most important result is Fig. 11. Here, the results of LR optimization for $T = 85$ are presented in declining order over 1000 conducted trials. Then, for each optimized LR value, we introduce the corresponding "expected" LR value of the true covariance matrix $\mathrm{LR}[\mathbf{T}_N]$ and the LR value of the initial Toeplitz matrix.

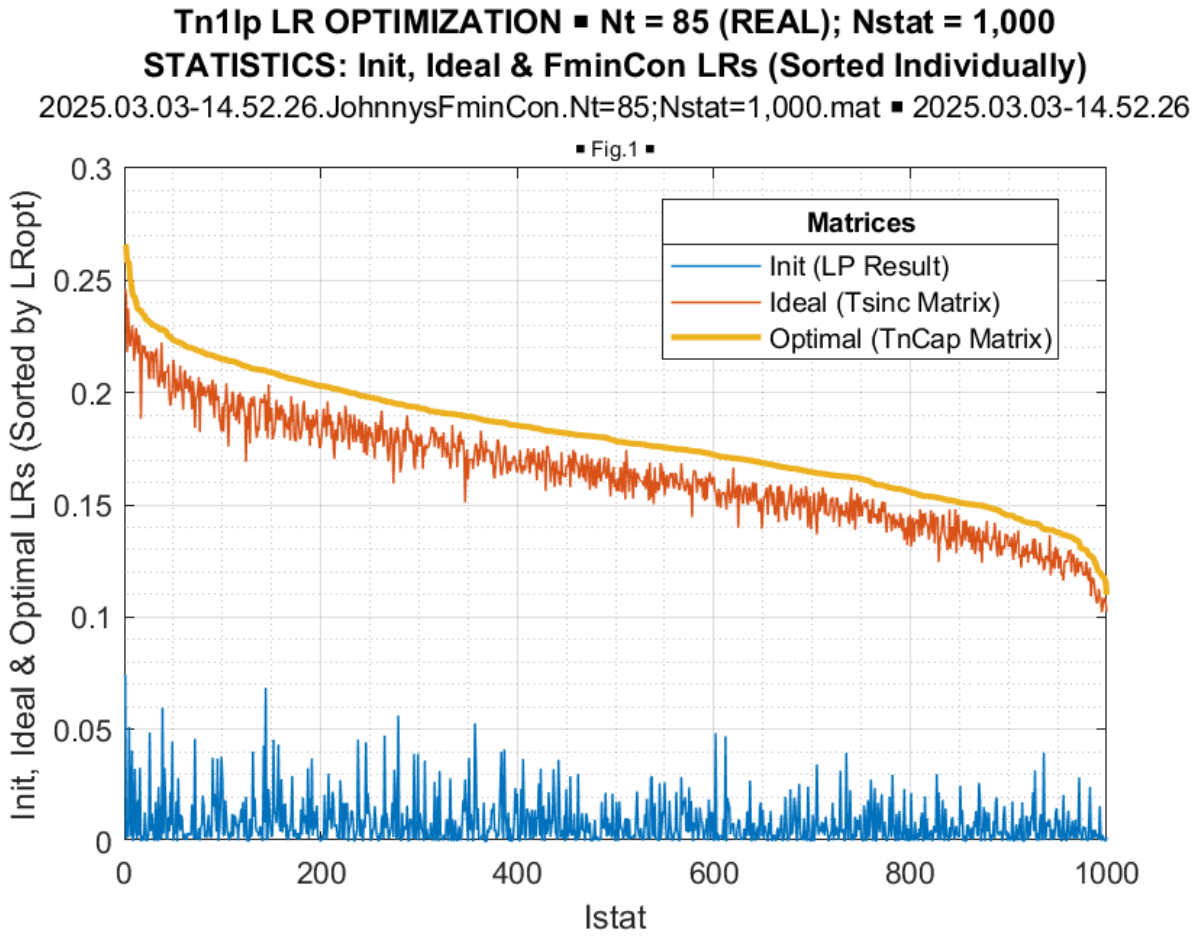

Fig. 11. LR optimization statistics, sorted by Optimal LR

In Fig. 12, we introduce values of these three likelihood ratios sorted in declining order individually.

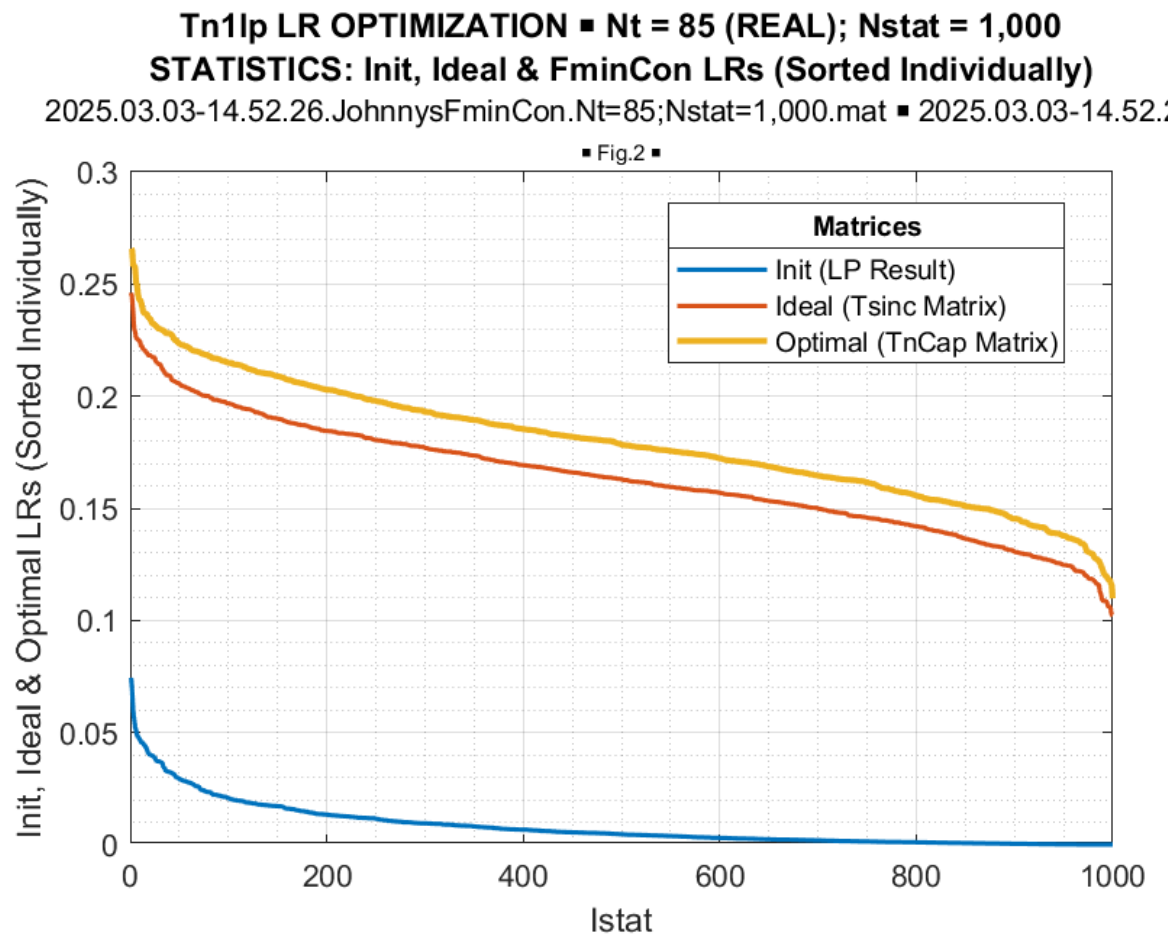

Fig. 12. LR optimization statistics, sorted individually

The most important "take away" from these figures is the greater than the "expected likelihood" results of the fmincon LR optimization in every conducted trial. The LR "gain," achieved by LR maximization concerning the "expected likelihood" $\mathrm{LR}[\mathbf{T}_N]$, is approximately the same for all trials. More details on this "gain" are provided by its distribution, presented in Fig. 13.

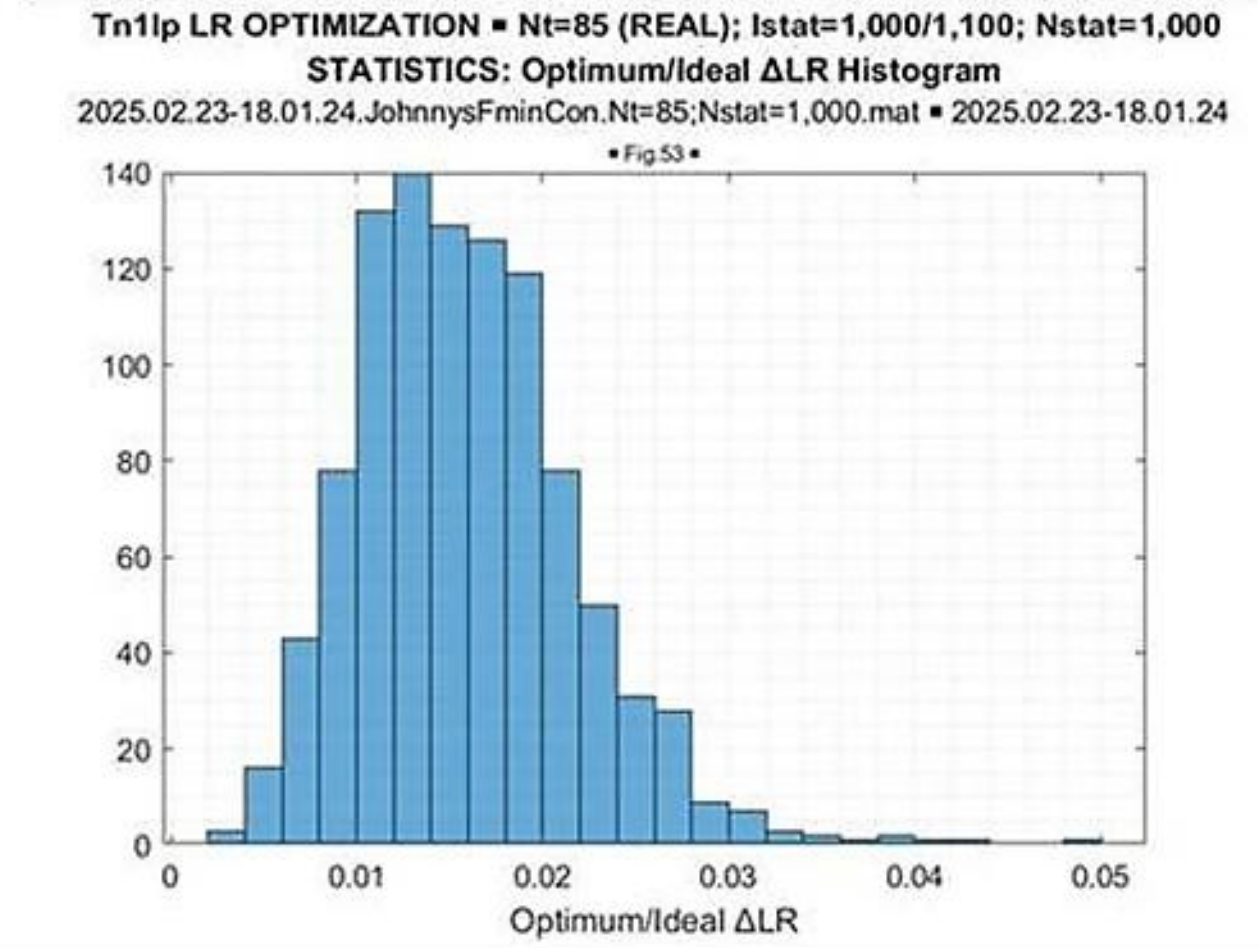

Fig. 13. ΔLR = (Optimal LR –Ideal LR)

These figures provide sufficient information on the nature of the fmincon convergence to the global LR maximum, which allows us to introduce the aggregated results.

In Fig. 14 and Fig. 15, we introduce the results of the likelihood ratio optimization as a function of the sample volume $T = 17 - 10^4$ and $T = 17 - 1000$, correspondingly.

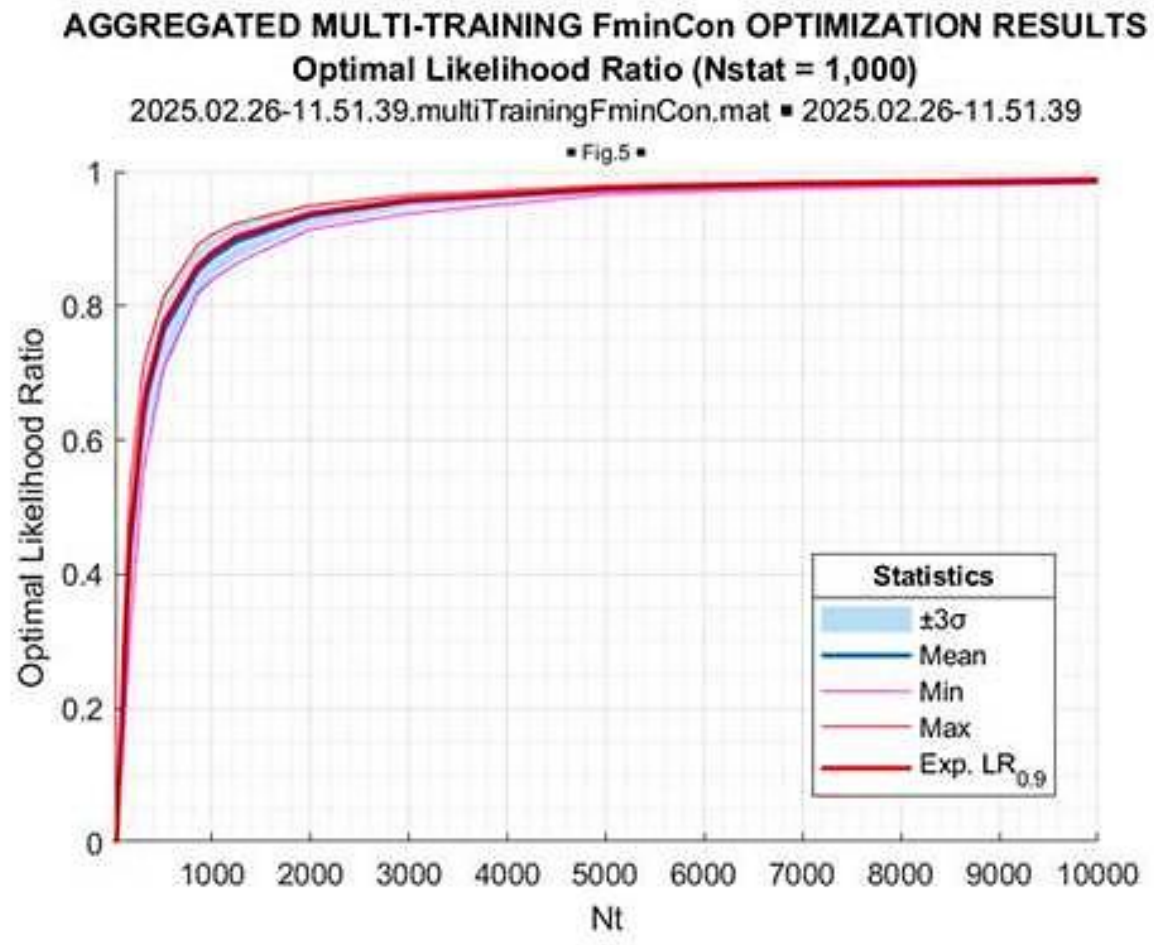

Fig. 14. Optimal LR

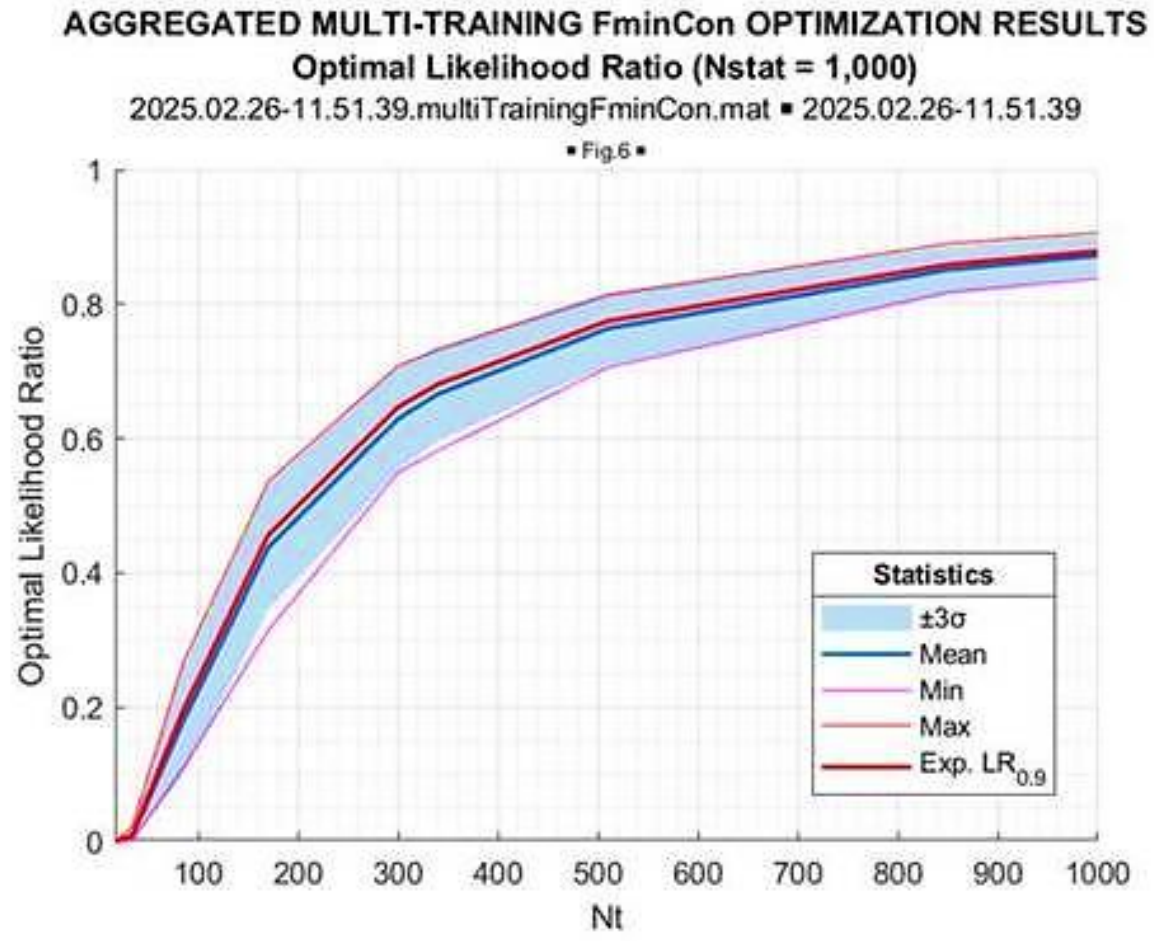

Fig. 15. Fmincon optimal LR

In each of these two figures, we introduce the mean (over $10^3$ trials), the minimum and maximum LR values for each sample volume, and the value of the "expected likelihood" $\mathrm{LR}[\mathbf{T}_N]$ at the probability level of 0.9. We also introduce the bounds for $\pm 3\sigma$ with respect to the mean optimized LR value. Fig. 14 demonstrates the sample volume range $T = 17 - 10^4$, while in Fig. 15 the sample range is reduced to $T = 17 - 10^3$.

For comparison, in Fig. 16 and Fig. 17, in the same format we introduce the results of the "expected likelihood" $\mathrm{LR}[\mathbf{T}_N]$ simulations.

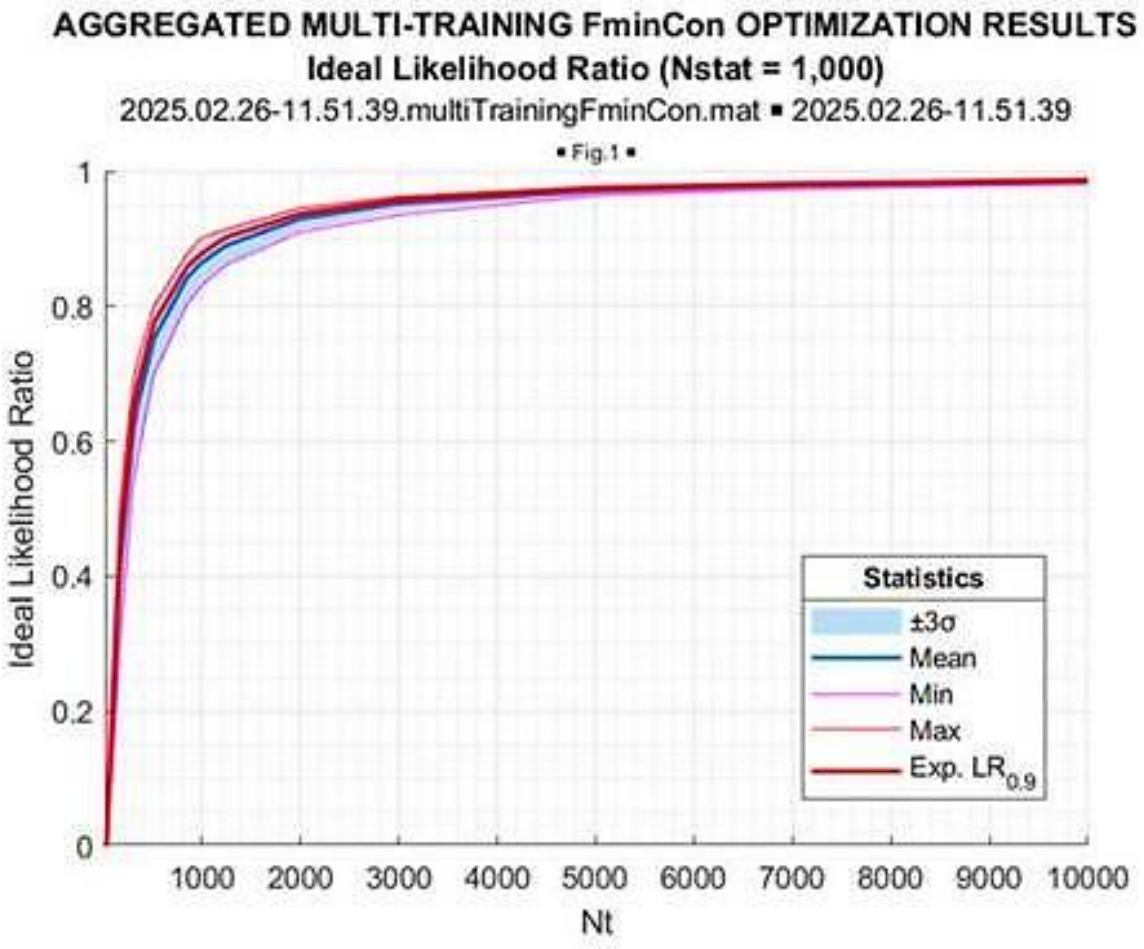

Fig. 16. Ideal LR

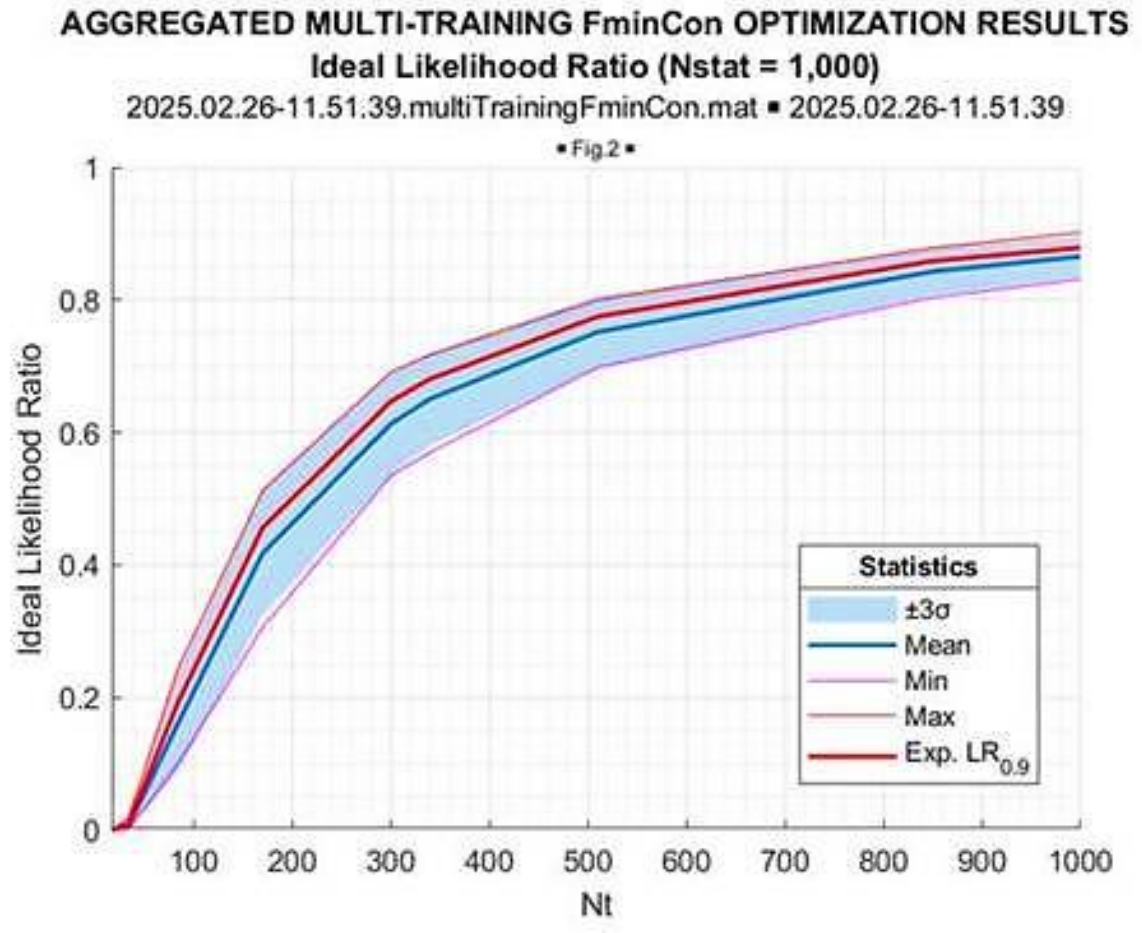

Fig. 17. Ideal LR

In Fig. 18 and Fig. 19, we present the statistical mean, maximum, and minimum values for the LR gain $(\mathrm{LR}[\mathbf{T}_{\mathrm{ML}}] - \mathrm{LR}[\mathbf{T}_N] = \Delta)$ of the optimum LR value $\mathrm{LR}[\mathbf{T}_{\mathrm{ML}}]$ with respect to the “expected likelihood” value, $\mathrm{LR}[\mathbf{T}_N]$ generated by the covariance matrix. The shadowed region corresponds to $0 < (\mathrm{LR}[\mathbf{T}_{\mathrm{ML}}] - \mathrm{LR}[\mathbf{T}_N]) \leq (\mathrm{mean} + 3\sigma)$.

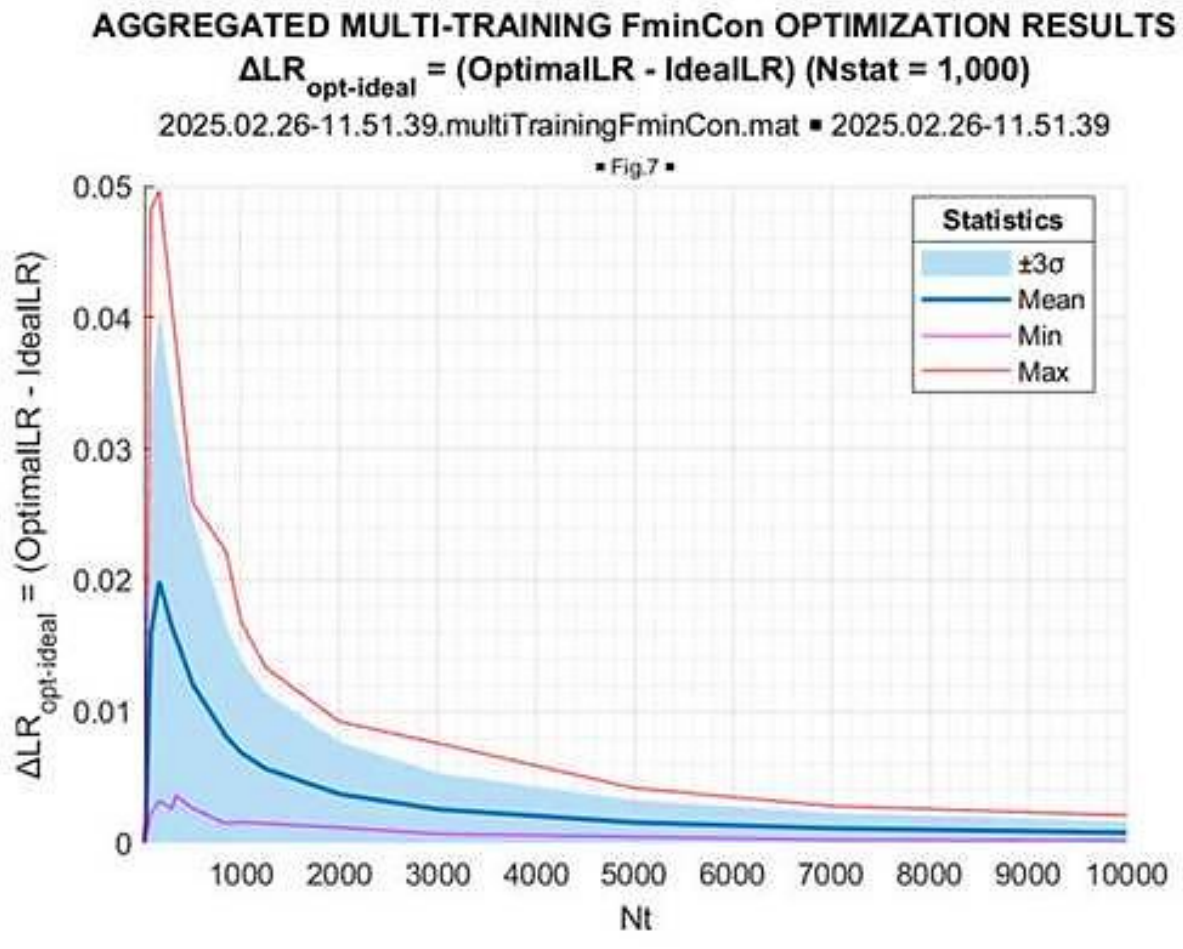

Fig. 18. ΔLR = (Optimal LR – Ideal LR)

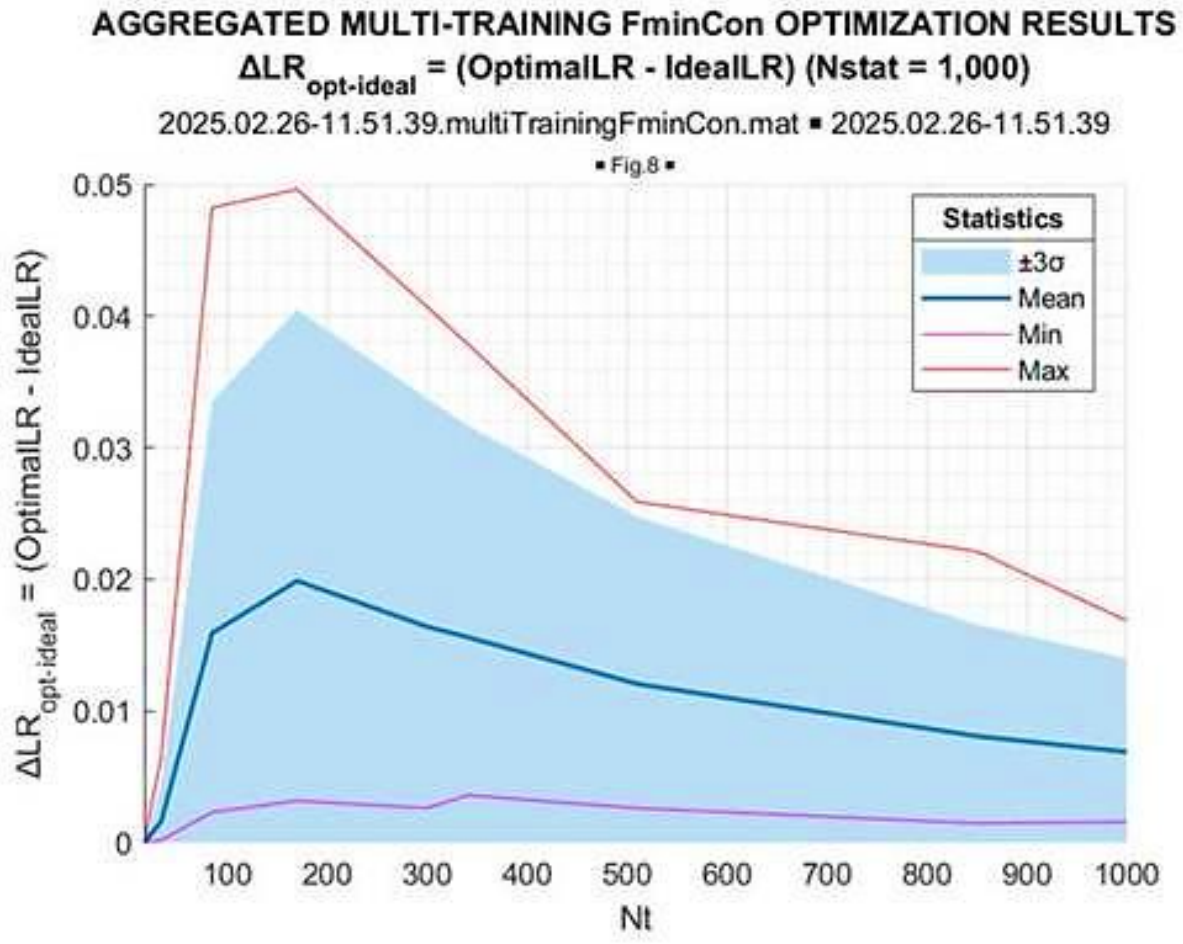

Fig. 19. ΔLR = (Optimal LR – Ideal LR)

Once again, we see that the optimized LR consistently exceeds the “expected likelihood” generated by the true covariance matrix $\mathrm{LR}[\mathbf{T}_N]$. This gain is maximal for the sample volume $\sim 10N(= 170)$, and it goes down for the larger sample volumes. More vividly, this LR gain is illustrated in Fig. 20, where the mean (over 1000 trials) LR gain with respect to “expected likelihood” $\mathrm{LR}[\mathbf{T}_N]$ is provided as the function of the training sample volume $T$.

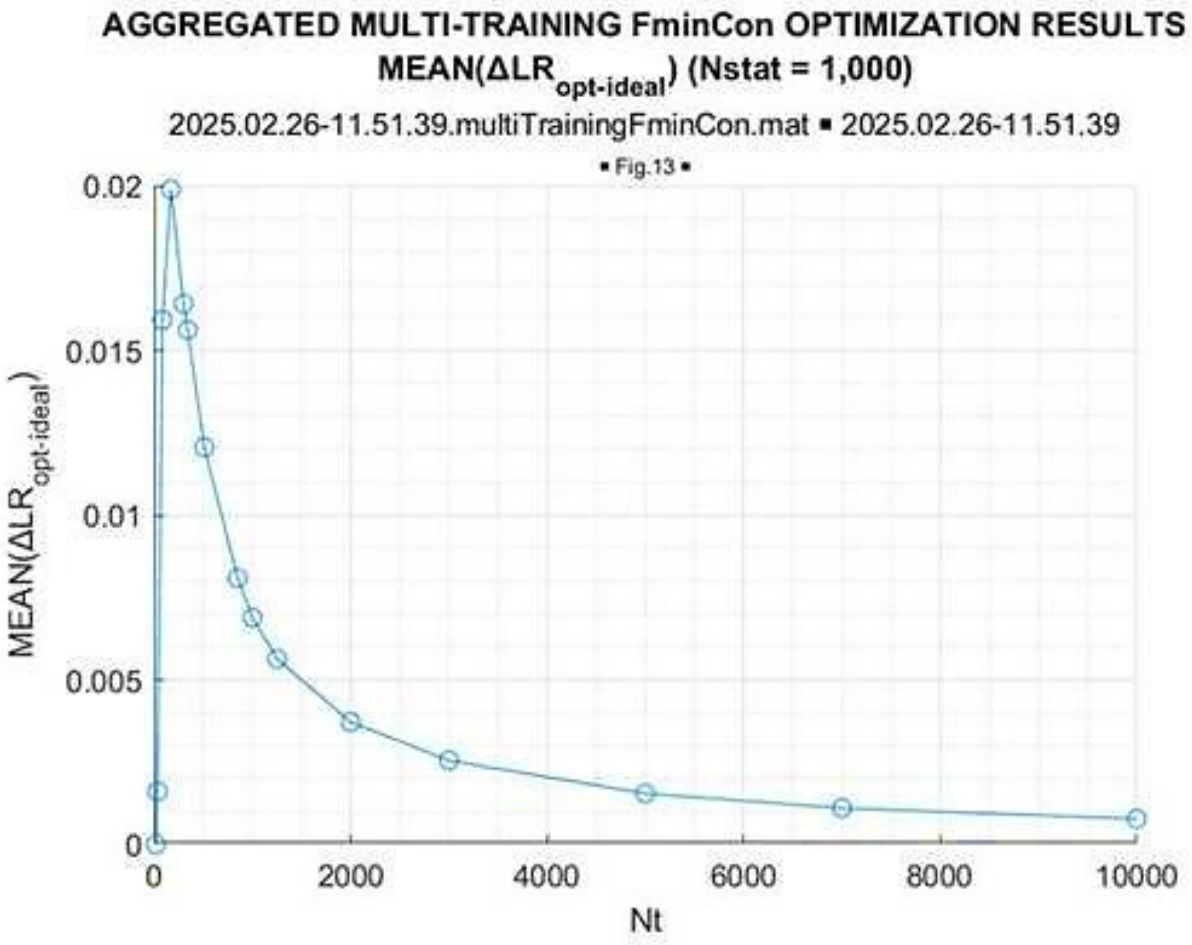

Fig. 20. Mean of ΔLR = (Optimal LR –Ideal LR)

Finally, we report on the comparison between the results of LR optimization within the class of the symmetric (real-valued) Toeplitz matrices the true matrix (10) - (11) belongs to and the results of optimization within the broader class of Hermitian Toeplitz matrices using the same symmetric Toeplitz matrix for initialization. The results of a specific example with $T = 85$ are illustrated in Fig. 21 and Fig. 22, for the real-valued and complex-valued Toeplitz matrices, correspondingly.

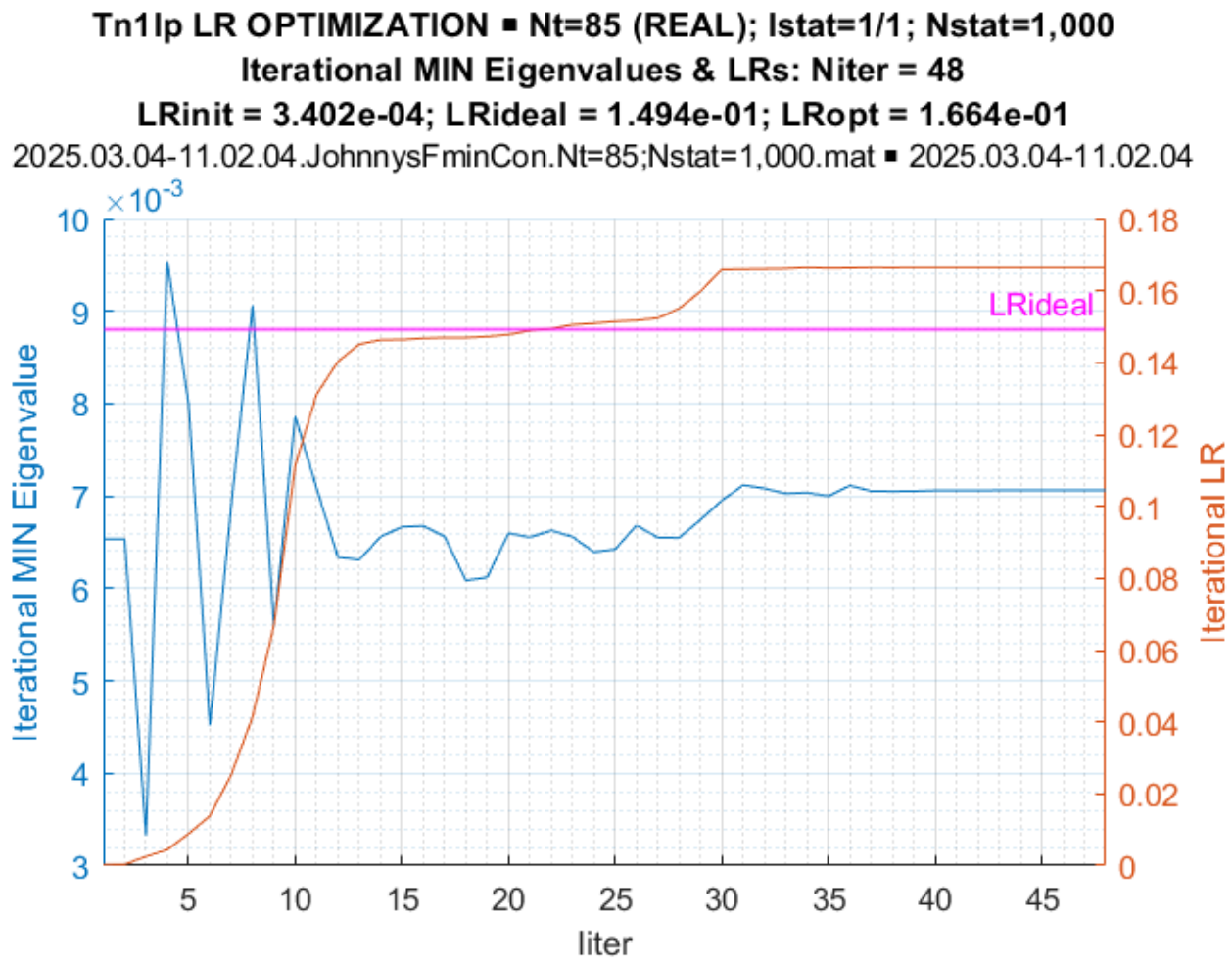

Fig. 21. Fmincon optimization: Iterative Min eigenvalues and LRs (real-valued Toeplitz matrices)

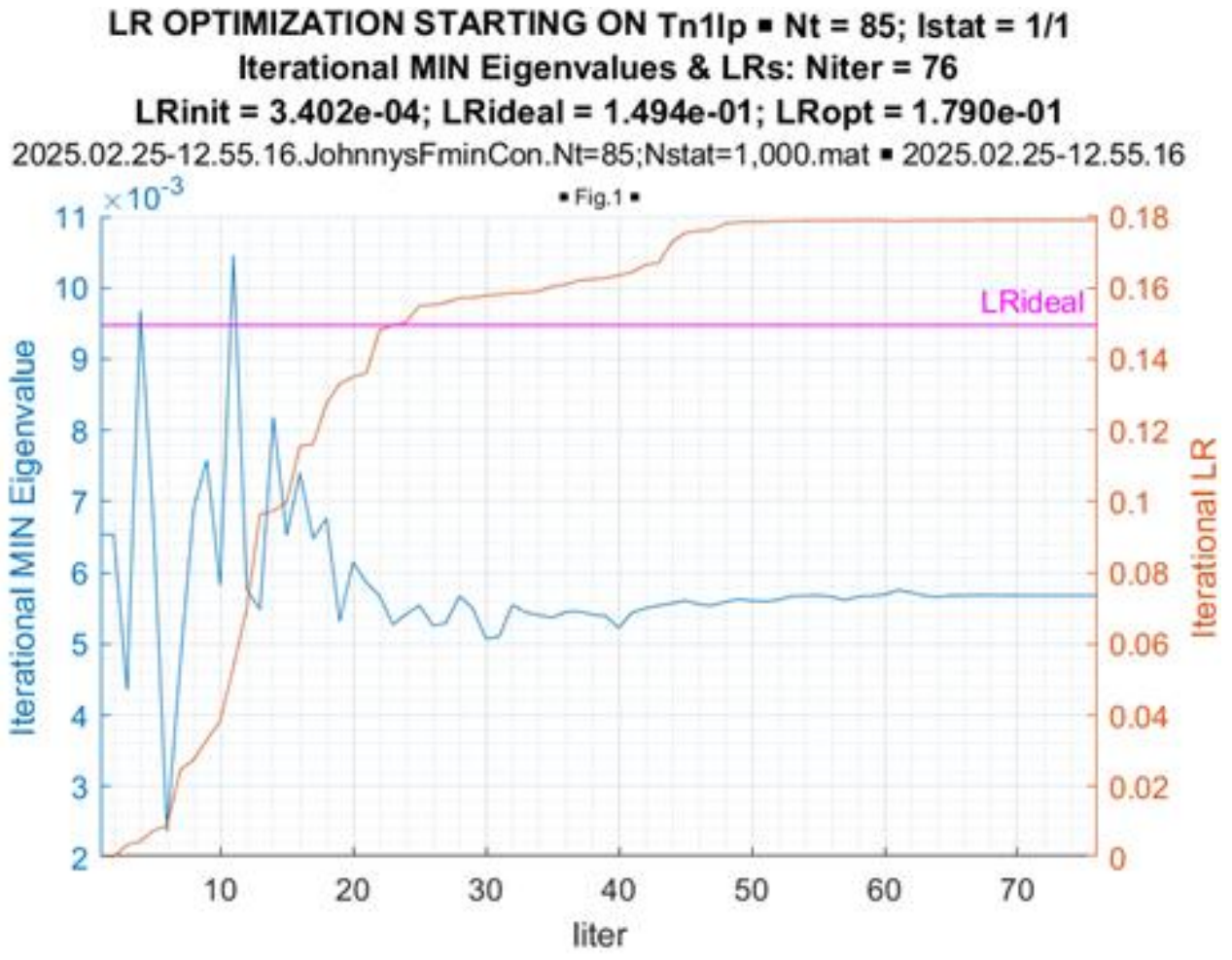

Fig. 22. Fmincon optimization: Iterative Min eigenvalues and LRs (complex-valued Toeplitz matrices)

For the symmetric Toeplitz matrix, the fmincon routine converged after ~10-th iteration to the level $\mathrm{LR}\left[\mathbf{T}_{\mathrm{ML}}^{\mathrm{R}}\right] = 0.1494$, while for the Hermitian Toeplitz matrix case, after ~75 iterations, the algorithm converged to the larger LR value $\mathrm{LR}\left[\mathbf{T}_{\mathrm{ML}}^{\mathrm{C}}\right] = 0.1790$.

More significant LR gain, achieved by LR optimization in the class of Hermitian Toeplitz matrices, is also evident from comparing the familiar Fig. 21 (real-valued case) with Fig. 22 (complex-valued case) data.

Sample pdf's of the LR gains provided by LR optimization concerning the $\mathrm{LR}[\mathbf{T}_N]$ true (symmetric) Toeplitz matrix for the symmetric Toeplitz matrix (Fig. 23) and for the Hermitian optimized Toeplitz matrix (Fig. 24), demonstrate that these LR gains are practically doubled for the optimization of Hermitian Toeplitz matrix, compared with the optimized symmetric Toeplitz matrix. Since the class of Hermitian Toeplitz matrices includes the symmetric matrices, this result is not surprising and once again supports our conclusion that the globally optimal results were achieved.

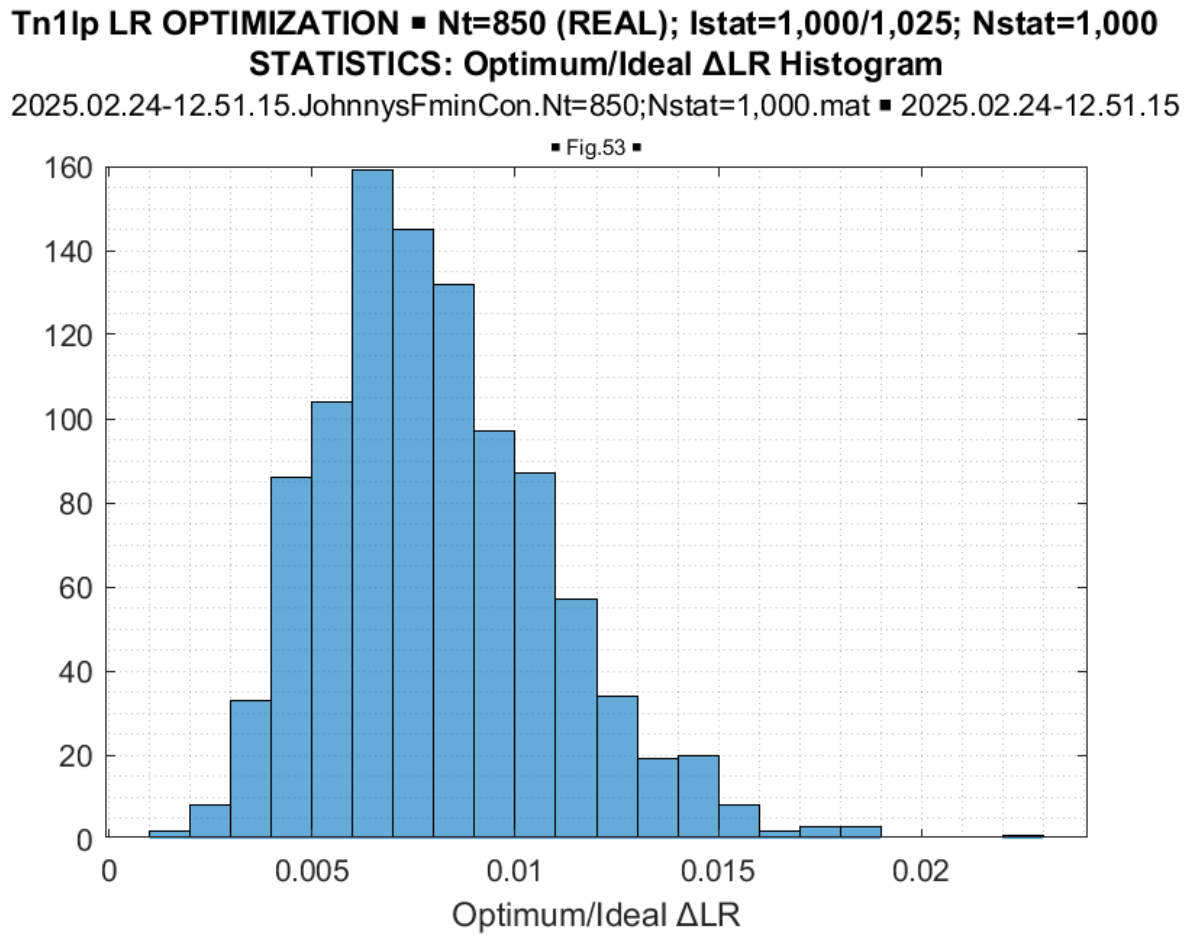

Fig. 23. Real-valued Toeplitz matrices:
ΔLR = (Optimal LR –Ideal LR) Histogram

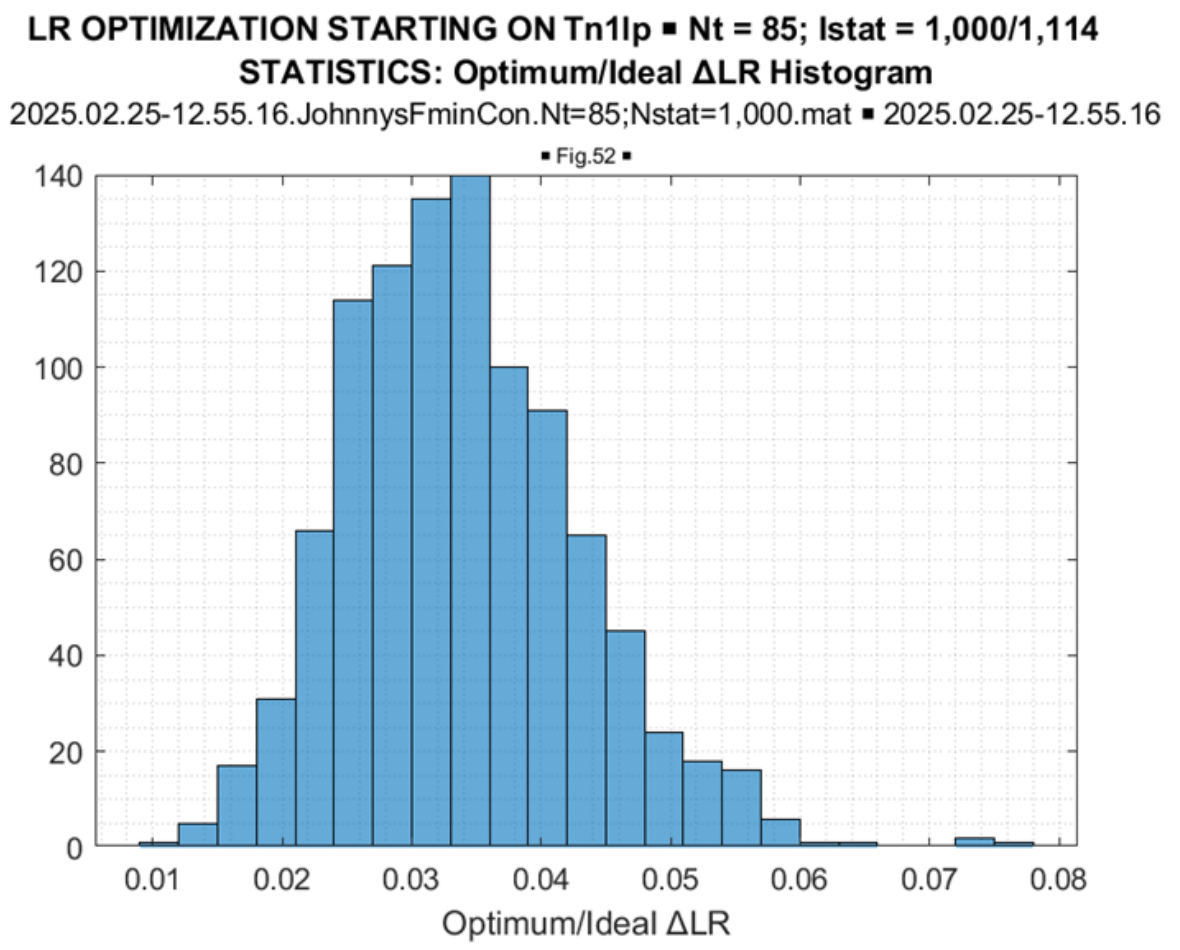

Fig. 24. Complex-valued Toeplitz matrices:
ΔLR = (Optimal LR –Ideal LR) Histogram

## VII. Conclusions and Recommendations

In this paper, we investigated the numerical methods for the symmetric (real-valued) Toeplitz covariance matrix estimation, focusing mainly on the methods available only for these matrices. The unique methodology is based on the M.T. Chu theorem, which suggests that the symmetric Toeplitz matrix of a ULA array, operating in the oversampled regime ($d/\lambda < 0.5$), is uniquely specified by the matrix elements moduli and eigenvalues. Since both sets of these parameters do not depend on the beam-steering phase progression and/or the phase "calibration" errors, this theorem suggests the unique possibility of the symmetric Toeplitz matrix reconstruction using the estimates of the sample covariance matrix elements' moduli and eigenvalues.

First, we confirmed that given the moduli of the elements and eigenvalues of a p.d. symmetric Toeplitz matrix, we can uniquely reconstruct this matrix. The true Toeplitz covariance matrix gets accurately reconstructed by the most simplistic "integer maximum element" algorithm. Since the true covariance matrix is usually not known a priori, in the most practical applications, we had to derive the estimates of the moduli and eigenvalues from the sample covariance Hermitian matrix, averaged over a finite number $T$ of the i.i.d. training samples.

We demonstrated that the "integer maximum element" algorithm, successfully operated on the accurate moduli and eigenvalues of this Toeplitz matrix, is inappropriate for operations with the estimates of these parameters drawn from the sample covariance matrix $\widehat{\mathbf{R}}_N$. More specifically, the i.i.d. training sample support T should be impractically large for this algorithm to work. For $N = 17$-element ULA and sample volume of $T = 17 \cdot 10^3$(!), the likelihood ratio of the solution is LR = 0.4, instead of the "expected likelihood" of the true matrix, which is equal to 0.98. For all "reasonable" training sample support volumes ($T = (2 - 10)N$), the redundancy averaged moduli of the sample matrix's diagonal elements do not often allow for a single positive definite Toeplitz matrix formation by selecting the appropriate sign changes over this matrix's sub- (and super-) diagonal elements.

This fact was established by testing all possible (65,535) sign combinations over the $N = 17$-element Toeplitz matrix sub- (and super-) diagonals for $T = 85$. This analysis demonstrated that the redundancy averaging over the moduli of the sub-diagonal elements of the sample matrix $\widehat{\mathbf{R}}_N$ is not a consistent estimate of the Toeplitz matrix lags. Since an appropriate alternative for the Toeplitz matrix moduli estimation does not exist, we had to develop an optimization sequence that overcame this limitation.

In Sec. III, we proposed several techniques with the initial integer optimization of the "best" position of the distributed sign inversion over the matrix diagonals, followed by "trimming" the averaged moduli to convert the matrix into a positive definite one. While our interest at this stage is in the potential efficiency of this symmetric matrix reconstruction, we applied the most computationally involved option where the LP conversion to the p.d. Toeplitz matrix accompanied each sign inversion testing. With this conversion, to select the "best" conversion at each probe, we compared the properties of the p.d. Toeplitz symmetric matrices.

Another serious problem is that the symmetric Toeplitz matrix reconstruction using the "moduli and eigenvalues" of the sample matrix is logical for applications when we cannot use the entire sample matrix $\widehat{\mathbf{R}}_N$ for optimization. The presence of the phase beam-steering progression and/or "calibration" phase errors are typical reasons for such a condition. Therefore, since we cannot use the likelihood ratio criterion at this stage, we had to apply the related to the maximum likelihood criteria.

We demonstrated that none of the considered alternative criteria is fully adequate to the maximum likelihood. We selected the best one, the minimax distance between the set of specified eigenvalues and eigenvalues of the reconstructed matrix. We provided the statistical analysis of this approach and compared the results with the impractical option with the clairvoyant knowledge of the sample matrix. Our analysis demonstrated that the reconstructed Toeplitz matrices have, on average, up to an order of magnitude worse LR compared with the LR of the true covariance matrix. For many practical applications, this accuracy of the reconstructed symmetric Toeplitz covariance matrix is sufficient, while the need to achieve the genuinely global LR maximum naturally remained.

The transition from "moduli and eigenvalues" symmetric reconstructed Toeplitz matrices, used as the initial solutions for the problem of ML Toeplitz matrix estimation, to the ML symmetric Toeplitz matrices, is already non-unique for the symmetric Toeplitz matrices. We used the MATLAB fmincon optimizer for this transition, using the derived "moduli and eigenvalues" solutions as the initial ones in this iterative algorithm. We also proposed an LR-based algorithm for the interim improvement of the "moduli and eigenvalues" Toeplitz matrices technique. Yet, for the considered problem, this interim step that had to increase the probability of getting the globally optimum solution was not required since all successful fmincon trials converged to the globally optimal solution. More specifically, in some rare cases, fmincon produced non-positive definite matrices, but all generated positive definite matrices converged to the global extremum. In these simulations, the global extremum was identified if its likelihood value exceeded the LR "expected likelihood" value of the true Toeplitz matrix $\mathbf{T}_N$, and if, starting from the true Toeplitz matrix $\mathbf{T}_N$, we converged to the same solution.

The authors recognize that the applied computational algorithms need to be upgraded to be considered for practical applications. Yet, they seem to be quite adequate for the potential maximum likelihood covariance matrix estimation efficiency analysis.